\documentclass{article}

\usepackage[utf8]{inputenc} %
\usepackage[T1]{fontenc}    %
\usepackage{hyperref}       %
\usepackage{url}            %
\usepackage{booktabs}       %
\usepackage{amsfonts}       %
\usepackage{nicefrac}       %
\usepackage{microtype}      %
\usepackage{xcolor}         %
\usepackage{xspace}
\usepackage{colortbl}
\usepackage{multirow}
\usepackage{graphicx} 
\usepackage{wrapfig}
\usepackage{subcaption}
\usepackage{adjustbox}
\usepackage{array}
\usepackage{tikz}
\usepackage{soul}
\usetikzlibrary{fit,positioning,calc}
\usepackage[T1]{fontenc}
\usepackage[frozencache=true,cachedir=minted-cache]{minted}
\usepackage[most,minted]{tcolorbox}
\usepackage{amsmath,amsfonts,bm}
\usepackage{amssymb}
\usepackage{mathtools}
\usepackage{amsthm}

\definecolor{strings}{rgb}{.824,.251,.259}
\definecolor{keywords}{rgb}{.224,.451,.686}
\definecolor{comment}{rgb}{.322,.451,.322}
\definecolor{lightcyan}{rgb}{0.88,1,1}
\definecolor{solidago}{RGB}{245,245,220}
\usepackage{pifont}
\newcommand\numberthis{\addtocounter{equation}{1}\tag{\theequation}}

\def\eqref#1{equation~\ref{#1}}

\def\1{\bm{1}}

\def\vtheta{{\bm{\theta}}}

\DeclareMathAlphabet{\mathsfit}{\encodingdefault}{\sfdefault}{m}{sl}
\SetMathAlphabet{\mathsfit}{bold}{\encodingdefault}{\sfdefault}{bx}{n}

\def\sS{{\mathbb{S}}}

\usepackage[accepted]{icml2024}

\newcommand{\lb}{\par\,\,\(\hookrightarrow\)\enspace}

\usepackage[capitalize,noabbrev]{cleveref}
\crefname{section}{§}{§§}
\Crefname{section}{§}{§§}
\crefformat{section}{#2§#1#3}
\crefmultiformat{section}{\S\S#2#1#3}{ and~#2#1#3}{, #2#1#3}{, and~#2#1#3}
\creflabelformat{equation}{#2\textup{#1}#3}

\definecolor{strings}{rgb}{.824,.251,.259}
\definecolor{keywords}{rgb}{.224,.451,.686}
\definecolor{comment}{rgb}{.322,.451,.322}
\definecolor{lightcyan}{rgb}{0.88,1,1}
\definecolor{solidago}{RGB}{245,245,220}
\definecolor{nicedarkblue}{RGB}{137,175,201}
\definecolor{pre}{RGB}{251,232,169}
\definecolor{mid}{RGB}{251,231,235}
\definecolor{suf}{RGB}{209,224,242}
\definecolor{nicegreen}{RGB}{54,171,18}
\definecolor{eot}{rgb}{0.88,1,1}
\definecolor{applegreen}{rgb}{0.55, 0.71, 0.0}
\definecolor{babyblue}{rgb}{0.54, 0.81, 0.94}
\definecolor{blizzardblue}{rgb}{0.67, 0.9, 0.93}
\definecolor{darkseagreen}{rgb}{0.56, 0.74, 0.56}
\definecolor{dollarbill}{rgb}{0.52, 0.73, 0.4}
\definecolor{arylideyellow}{rgb}{0.91, 0.84, 0.42}
\definecolor{bananamania}{rgb}{0.98, 0.91, 0.71}
\definecolor{bluegray}{rgb}{0.4, 0.6, 0.7}
\definecolor{buff}{rgb}{0.94, 0.86, 0.51}
\definecolor{deepsaffron}{rgb}{1.0, 0.6, 0.2}
\definecolor{lava}{rgb}{0.57, 0.01, 0.13}

\newcommand{\dsonek}{\textsc{DS-1000}\xspace}

\newcommand{\mbpp}{\textsc{MBPP}\xspace}

\newcommand{\gsm}{\textsc{GSM8K}\xspace}
\newcommand{\multiple}{\textsc{MultiPL-E}\xspace}
\newcommand{\humaneval}{\textsc{HumanEval}\xspace}
\newcommand{\humanevalplus}{\textsc{HumanEval}\textsuperscript{+}\xspace}

\newcommand{\pprefix}{\texttt{prefix}\xspace}
\newcommand{\psuffix}{\texttt{suffix}\xspace}
\newcommand{\pmiddle}{\texttt{middle}\xspace}
\newcommand{\psp}{\texttt{suffix\textsuperscript{p}}\xspace}
\newcommand{\psc}{\texttt{suffix\textsuperscript{c}}\xspace}

\newcommand{\codellama}{\textsc{Code~Llama}\xspace}
\newcommand{\codellamainst}{\textsc{Code~Llama~-~Instruct}\xspace}
\newcommand{\codellamapython}{\textsc{Code~Llama~-~Python}\xspace}
\newcommand{\starcoder}{\textsc{StarCoder}\xspace}
\newcommand{\starcoderbase}{\textsc{StarCoderBase}\xspace}

\newcommand{\deepseekcoder}{DeepSeek-Coder-Base\xspace}
\newcommand{\deepseekcoderinst}{DeepSeek-Coder-Instruct\xspace}
\newcommand\extrafootertext[1]{%
    \bgroup
    \renewcommand\thefootnote{\fnsymbol{footnote}}%
    \renewcommand\thempfootnote{\fnsymbol{mpfootnote}}%
    \footnotetext[0]{#1}%
    \egroup
}

\newcommand{\codecolorbox}[2]{%
  \begingroup\setlength{\fboxsep}{0pt}%
  \colorbox{#1}{#2}%
  \endgroup
}

\newenvironment{itemizesquish}{\begin{list}{\labelitemi}{\setlength{\itemsep}{-0.1em}\setlength{\labelwidth}{0.5em}\setlength{\leftmargin}{\labelwidth}\addtolength{\leftmargin}{\labelsep}}}{\end{list}}

\newtcolorbox[number format=\arabic]{DSOuterBox}[1]{
  size=title,
  arc=1.2mm,
  enhanced jigsaw,
  colframe=white!30!black,
  coltitle=white,
  boxrule=0.5mm,
  colback=white,
  coltext=black,
  title=#1,
}
\newtcolorbox[number format=\arabic]{DSInnerBox}[1]{
  size=title,
  arc=1.2mm,
  breakable,
  enhanced jigsaw,
  colframe=brown,
  coltitle=white,
  boxrule=0.5mm,
  colback=white,
  coltext=black,
  title=#1,
  sidebyside,
  sidebyside gap=6mm,
  sidebyside align=top seam,
}

\newtcolorbox[number format=\arabic]{twoColumnCode}[1]{
  size=title,
  arc=1.2mm,
  breakable,
  enhanced jigsaw,
  colframe=brown,
  coltitle=white,
  boxrule=0.5mm,
  colback=white,
  coltext=black,
  title=#1,
  sidebyside,
  sidebyside gap=6mm,
  sidebyside align=top seam
}
\newtcolorbox[number format=\arabic]{singleColumnCode}[1]{
  size=title,
  arc=1.2mm,
  breakable,
  enhanced jigsaw,
  colframe=red!40!black,
  coltitle=white,
  boxrule=0.5mm,
  colback=white,
  coltext=black,
  title=#1,
}
\newtcblisting{codelisting}[1]{
  listing engine=minted,
  minted language=#1,
  minted options={
    style=vs,
    escapeinside=||,
    breaklines,
    breakautoindent=false,
    breaksymbolleft=,
    breaksymbolindentleft=0pt,
    breaksymbolsepleft=0pt,
    fontsize=\scriptsize
  },
  listing only,
  size=title,
  breakable,
  enhanced jigsaw,
  frame hidden,
  boxrule=0pt,
  colback=white,
  coltext=black,
}

\icmltitlerunning{Self-Infilling Code Generation}

\begin{document}

\twocolumn[
\icmltitle{Self-Infilling Code Generation}

\icmlsetsymbol{equal}{*}

\begin{icmlauthorlist}
\icmlauthor{Lin Zheng}{sch,equal}
\icmlauthor{Jianbo Yuan}{comp}
\icmlauthor{Zhi Zhang}{comp}
\icmlauthor{Hongxia Yang}{comp}
\icmlauthor{Lingpeng Kong}{sch}
\end{icmlauthorlist}

\icmlaffiliation{comp}{ByteDance Inc}
\icmlaffiliation{sch}{The University of Hong Kong}

\icmlcorrespondingauthor{Lin Zheng}{lzheng2@cs.hku.hk}

\icmlkeywords{Machine Learning, Code Generation, Language Models}

\vskip 0.3in
]

\printAffiliationsAndNotice{\textsuperscript{*}This work was done during an internship at ByteDance.}

\begin{abstract}

In this work, we introduce self-infilling code generation, a general framework that incorporates infilling operations into auto-regressive decoding.
Our approach capitalizes on the observation that recent infilling-capable code language models can perform \emph{self-infilling}: whereas conventional infilling is designed to fill in the middle based on a predefined prefix and suffix, self-infilling sequentially generates both such surrounding context and the infilled content.
We utilize self-infilling to introduce novel interruption and looping mechanisms in conventional decoding, evolving it into a non-monotonic process.
Interruptions allow for postponing the generation of specific code until a definitive suffix is established, enhancing control during decoding.
Meanwhile, the looping mechanism, which leverages the complementary nature of self-infilling and left-to-right decoding, can iteratively update and synchronize each piece of generation cyclically.
Extensive experiments across a variety of code generation benchmarks demonstrate that decoding with self-infilling not only improves the output quality but also regularizes the overall generation, which effectively mitigates potential degeneration and scaffolds code to be more consistent with intended functionality.

\end{abstract}

\section{Introduction}
\label{sec:intro}
Contemporary large language models have achieved excellent performance in tasks related to code generation and understanding \citep{austin2021program,chen2021codex,fried2023incoder,nijkamp2023codegen,li2022alphacode,anil2023palm2,openai2023gpt4,touvron2023llama,li2023starcoder,touvron2023llamav2,muennighoff2023octopack,nijkamp2023codegen2,roziere2023codellama,xie2023text2reward,xu2024lemur}. Building upon this success, an active line of research aims to endow these language models with enhanced free-form generation capacities \citep{fried2023incoder,bavarian2022fim,allal2023santacoder,nijkamp2023codegen2,li2023starcoder,roziere2023codellama}, where models learn to \emph{infill} content considering both preceding and subsequent contexts. Such capabilities are instrumental for numerous downstream code-related tasks that require a bidirectional context, including but not limited to partial code completion, docstring generation, and type prediction \citep{fried2023incoder,li2023starcoder,roziere2023codellama}.

Established practices for fostering infilling capabilities in language models involve training with an explicit infilling objective \citep{raffel2020t5,bavarian2022fim,tay2023ul2,anil2023palm2,fried2023incoder,nijkamp2023codegen2}, which directs the model to predict the missing span of sequence tokens given the surrounding context. Despite a substantial allocation of computational resources dedicated to the specialized infilling training \citep{bavarian2022fim,li2023starcoder,roziere2023codellama}, most code generation systems still persist in a strictly left-to-right fashion. It remains unclear how the acquired proficiency in infilling could aid in synthesizing \emph{complete} code beyond narrowly tailored partial infilling tasks \citep{bavarian2022fim,fried2023incoder,allal2023santacoder}.

In this work, we investigate the integration of infilling capabilities into the generation process of code language models. We start with a seemingly straightforward yet overlooked observation: recent open-source code language models, specifically those trained with fill-in-the-middle objectives \citep{bavarian2022fim} (e.g., \starcoder \citep{li2023starcoder} and \codellama \citep{roziere2023codellama}), inherently possess \emph{self-infilling} capabilities (\cref{model:self_infilling}). Unlike regular infilling operations that demand a partial surrounding context as input, self-infilling autonomously generates \emph{both} the surrounding context and the infilled content.

\begin{figure}[t]
\centering
\includegraphics[width=0.99\columnwidth]{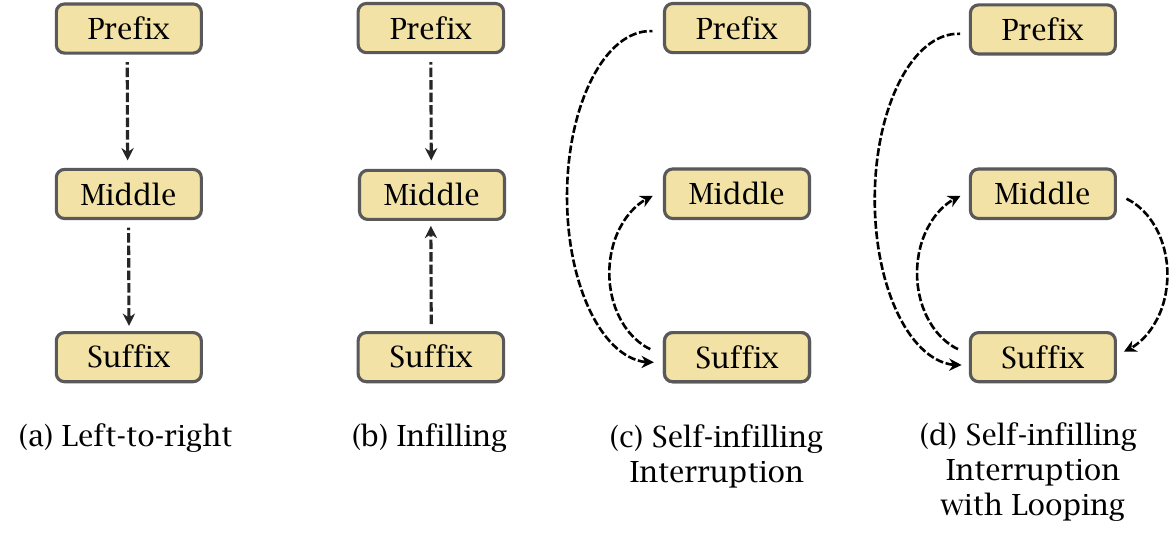}
\caption{Schematic illustrations of various decoding approaches for code generation. \textbf{(a)} and \textbf{(b)} represent standard left-to-right decoding and infilling operations, respectively. Whereas infilling requires the \emph{user-provided} prefix and suffix, self-infilling interruption \textbf{(c)} autonomously generates these segments. \textbf{(d)} further expands on self-infilling by incorporating a looping mechanism.}
\label{fig:self-infill}
\end{figure}

We leverage this finding to develop an \emph{interruption} mechanism (\cref{model:self_infilling:interruption}) in conventional left-to-right decoding, transforming it into a non-monotonic process. This allows the language model to temporarily halt the decoding process when necessary, craft a suffix, and then return to the interruption point to infill the bypassed context (\cref{fig:self-infill}c). To handle the variable length of the skipped context and enforce an appropriate suffix, we introduce \emph{suffix prompting} by initiating the suffix with a prespecified yet common prompt. The interruption mechanism is particularly useful in high-entropy situations, where conventional decoding methods falter due to uncertainty in predicting the next token and might cause propagation of errors through the context, known as the exposure bias \citep{ranzato2015exposurebias,bengio2015scheduledsampling}. Instead, self-infilling interruption offers an \emph{easy-first} alternative by deferring the generation of more difficult tokens and crafting a well-definitive suffix. This suffix works as an anchor to scaffold the code and produce a more structured context for the subsequent infilling, which ensures the final generation is logically consistent with the intended functionality and enjoys improved regularity.

Besides, self-infilling implies a \emph{looping} mechanism (\cref{model:self_infilling:looped}) to further enhance the decoding process. Leveraging the complementary nature of left-to-right and self-infilling methods, it iteratively updates code snippets with broader contexts, wherein the output from one mode seamlessly informs the input to the other (\cref{fig:self-infill}d). Specifically, the suffix in self-infilling is generated based solely on the prefix with a rather limited context. This can be addressed by chaining a subsequent left-to-right decoding phase, enabling the suffix to be re-generated within an expanded context that includes the infilled middle. Similarly, we can continue by appending the next self-infilling to synchronize the middle and the prefix based on the latest suffix information. Such piece-wise synchronization allows each snippet to be recurrently updated with more informative contexts.

Intuitively, the developed interruption and looping mechanism mirrors human coding practices, which typically do not conform strictly to a linear, left-to-right progression. Instead, code is usually written via a dynamic process of continuous refinement, editing, and reconnection of fragments. While the replication of human coding strategies may not necessarily be the optimal approach for language models, our extensive experiments (\cref{exp}) on various code benchmarks verify the effectiveness of self-infilling generation. We demonstrate that our framework brings significant improvements in the quality and regularity of code generation compared to conventional left-to-right approaches.

\section{Self-infilling Code Generation}
\label{model}
In this section, we introduce self-infilling, the built-in capability of models trained with \emph{fill-in-the-middle} \citep[\cref{model:self_infilling};][]{bavarian2022fim}. We then detail our integration of self-infilling into the decoding process, including the developed interruption (\cref{model:self_infilling:interruption}) and looping (\cref{model:self_infilling:looped}) mechanisms.

\subsection{FIM Training Entails Self-infilling}
\label{model:self_infilling}

\paragraph{Fill-in-the-middle (FIM) Training.} Throughout this work, we focus on code language models with causal decoder-only Transformers \citep{vaswani2017attention,radford2018gpt}, which are currently the dominant paradigm at large scale. Most of these language models are trained with conventional left-to-right next token prediction \citep{radford2018gpt,radford2019gpt2}. Despite compelling performance \citep{austin2021program,chen2021codex,nijkamp2023codegen}, such training objective only permits left-to-right generation for downstream applications and restricts more versatile generation tasks like infilling, which necessitates a non-causal (or bidirectional) context. A recent practice to address this limitation is training with the \emph{Fill-In-the-Middle} (FIM) objective \citep{bavarian2022fim}, which is extensively employed in recent code models \citep{allal2023santacoder,li2023starcoder,roziere2023codellama}. FIM randomly splits input raw code into three pieces (\pprefix, \pmiddle, \psuffix) and rearranges them to form a permuted sequence \texttt{[<PRE>;prefix;<SUF>;suffix;<MID>; middle;<EOT>]}, where \texttt{;} denotes concatenation and particular sentinel tokens are interleaved to mark the boundary of each piece.\footnote{An alternative format, known as the SPM mode \citep{bavarian2022fim}, structures the pieces as {\footnotesize\texttt{[<PRE>;<SUF>; suffix;<MID>;prefix;middle;<EOT>]}}. However, this permutation is not as useful in our setting, because we would like the suffix to be conditioned on the prefix context.} The model is then optimized to maximize the factorized likelihood $p(\texttt{prefix})p(\texttt{suffix}\mid \texttt{prefix}) p(\texttt{middle}\mid \texttt{prefix}, \texttt{suffix})$ with each modeled in a standard left-to-right auto-regressive manner.

\paragraph{FIM Entails Self-infilling.}

Our key observation is that FIM \citep{bavarian2022fim} trains the model to predict the ``next'' token irrespective of the specific pieces (i.e., \pprefix, \pmiddle, or \psuffix) these tokens belong to. Consequently, the model learns not only to predict the \pmiddle conditioned on both \pprefix and \psuffix, but also to predict \psuffix only based on \pprefix. While the former is extensively studied in the literature as infilling \citep{donahue-etal-2020-enabling,bavarian2022fim,du2022glm,aghajanyan2022cm3,tay2022upalm,tay2023ul2,fried2023incoder}, the latter is rarely explored, which characterizes the distribution of \psuffix given \pprefix marginalized over possible outcomes of \pmiddle. This leads to \emph{self-infilling}, where the model can first generate a \psuffix based on \pprefix and then return to fill in the \pmiddle given the \emph{self-generated} context. We leverage this aspect to develop a non-monotonic decoding process, as detailed subsequently.

\subsection{Self-infilling Interruption}
\label{model:self_infilling:interruption}
Self-infilling introduces a dynamic decoding process with the \emph{interruption} mechanism. Specifically, given an initial input prompt denoted as \pprefix, the language model performs next-token prediction similar to regular decoding,
\begin{align*}
    \pprefix \sim p(\pprefix). \numberthis\label{eqn:si:prefix}
\end{align*}
Our approach enhances decoding by incorporating \emph{interruptions}. On invocation, the current decoding is suspended to generate a \psuffix, followed by filling in the \pmiddle:
\begin{align*}
    \psuffix &\sim p(\psuffix \mid \pprefix), \numberthis\label{eqn:si:suffix:no_sp}\\
    \pmiddle &\sim p(\pmiddle \mid \pprefix, \psuffix). \numberthis\label{eqn:si:middle}
\end{align*}
In general, interruptions can be triggered by various indicators, such as low likelihood or the occurrence of specific tokens. We implement a simple heuristic to signal interruptions by employing a probability threshold $\tau$. If the maximum probability over the next token falls below $\tau$, indicating high uncertainty, self-infilling is invoked. Contrasting with left-to-right decoding, self-infilling interruption adopts an \emph{easy-first} methodology and dynamically defers the generation of potentially difficult snippets. It is thus useful to mitigate divergent issues due to producing an error-prone context, known as the exposure bias \citep{bengio2015scheduledsampling,ranzato2015exposurebias}.

\paragraph{Suffix Prompting.}
Accurately crafting a proper \psuffix from only \pprefix is often challenging due to the indeterminate nature of the skipped segment \pmiddle during infilling pretraining. To guide suffix generation, we propose \emph{suffix prompting} that enforces \psuffix to start with specific tokens. For instance, in Python function generation \citep{chen2021codex,austin2021program}, a common ending is a return statement containing the keyword \mintinline{python}{return}. Such common keywords can be used as the \emph{suffix prompt} (denoted as \psp) to shape \psuffix.

Technically, we represent the whole input sequence as a quadruple in the following form, 
\begin{equation}
    [\pprefix;\,\, \pmiddle;\,\, \psp;\,\, \psc],
\label{eqn:self-infill-output-parse}
\end{equation}
where $\psuffix \coloneqq [\psp;\psc]$ and \psc is the \emph{suffix completion} of \psp. Given \pprefix and a predefined \psp, we can further expand upon \cref{eqn:si:suffix:no_sp} as follows,
\begin{align*}
    \numberthis\label{eqn:si:suffix:sp}
    \begin{split}
        \psc &\sim p(\psc \mid \pprefix, \psp), \\
        \psuffix &\coloneqq [\psp;\psc]. \\
    \end{split}
\end{align*}
The use of suffix prompting helps generate an appropriate \psuffix, which serves as a contextual anchor to scaffold the overall generation and ensure the output structure is logically consistent with the intended functionality.

In practice, self-infilling interruption is implemented through the manipulation of sentinel tokens. For instance, the language model is programmed to produce \texttt{<SUF>} as the next token upon interruption activation. Our detailed implementation is outlined in \cref{alg:self-infill} (\cref{app:pseudocode}).

\subsection{Decoding through a Looping Mechanism}
\label{model:self_infilling:looped}
In this section, we introduce a \emph{looping mechanism} to improve decoding, which interweaves self-infilling with left-to-right conditional generation to recurrently update snippets. Note that during self-infilling, \psuffix is generated based solely on \pprefix, a narrower context that excludes \pmiddle. This can be enhanced by chaining a subsequent left-to-right decoding phase to re-generate $\psuffix$ as
\begin{align*}
    \psp &\!\sim\! p(\psp \!\mid\! \pprefix, \pmiddle), \\
    \psc &\!\sim\! p(\psc \!\mid\! \pprefix, \pmiddle, \psp),
\end{align*}
allowing $\psuffix \coloneqq [\psp;\psc]$ to integrate information from both \pprefix and \pmiddle. Notably, the suffix prompt \psp is also updated through looping to become fully contextual instead of being specified \emph{a priori}.

This looping procedure can be continued to synchronize \pmiddle with the latest suffix information through the next self-infilling call. Instead of restarting self-infilling with \pprefix generation (\cref{eqn:si:prefix}), which would ignore the information from the last left-to-right decoding and repeat the first iteration, hereafter self-infilling begins with \cref{eqn:si:suffix:sp} to directly incorporate \psp.\footnote{It is also feasible to pass the entire generated \psuffix to initiate self-infilling with \pmiddle generation (\cref{eqn:si:middle}); however, this approach often yields inferior results, as shown in \cref{tb:ablation:suffix_split} and discussed in \cref{exp:ablation}.} To update the generated tokens of \pprefix (\cref{eqn:si:prefix}), we prepend them into \texttt{middle} and reset \pprefix to the original fixed input. This design choice corroborates prior research \citep{bavarian2022fim}, suggesting that while the infilled \pmiddle sometimes struggles to join \psuffix, it could adeptly continue \pprefix. During looping, transferring the output from self-infilling to left-to-right decoding is straightforward due to explicit sentinel tokens. However, to continue the looping mechanism from the left-to-right decoding phase, we have to parse the output into the quadruple (\pprefix, \pmiddle, \psuffix, \psp, \psc), as these segments are not explicitly delineated in left-to-right decodes. We explore several heuristic approaches for this segmentation in Function~\ref{fig:parse-function} and discuss them further in \cref{exp:ablation}.

As outlined in \cref{alg:loop}, the looping mechanism cyclically updates \pmiddle and \psuffix through self-infilling and left-to-right generation, respectively. Since each piece of the context is updated \emph{in situ}, this looping process, akin to a rolling window over the context or piece-wise Gibbs sampling, leads to continuous synchronization of each section. In addition, this mechanism allows all tokens to be conditioned on a more informative bidirectional context, overcoming the causal limitations of traditional decoding.

\begin{algorithm}[t] %
   \caption{Looping Mechanism}
   \label{alg:loop}
    \begin{algorithmic}
    \State {\bfseries Input:} $\texttt{prompt}$, the language model, suffix prompt tokens $\psp$, and number of iterations $N$.
    \State {\bfseries Output:} Generated code $y$.

    \State
    \State Set $\pprefix \gets \texttt{prompt}$ and $x \gets [\texttt{<PRE>};\pprefix]$;
    \State
    \For{$n = 1,2,\dots,N$}
        \State Invoke self-infilling generation (\cref{alg:self-infill}) with\lb input $x$ to output $x'$;
        \State Parse $x'$ into\lb $(\pprefix', \pmiddle', \psp', \psc')$;
        \For{$\texttt{p} \in \scalebox{0.92}[0.92]{(\pprefix, \pmiddle, \psp, \psc)}$}
            \State \Comment{Update each piece to its latest version.}
            \State Update $\texttt{p} \gets \texttt{p}'$;
        \EndFor
        \State
        \State Set $x \gets [\pprefix;\pmiddle]$;
        \State
        \State Invoke left-to-right generation (\cref{alg:left-to-right}) with\lb input $x$ to output $x'$;
        \State Parse $x'$ via \texttt{l2r\_parser()} (Function~\ref{fig:parse-function}) into\lb $(\pprefix', \pmiddle', \psp', \psc')$;
        \For{$\texttt{p} \in \scalebox{0.92}[0.92]{(\pprefix, \pmiddle, \psp, \psc)}$}
            \State Update $\texttt{p} \gets \texttt{p}'$;
        \EndFor
        \State
        \If{$n \neq N$}
            \State \Comment{Construct new input $x$ for the next cycle.}
            \State $x \gets [\texttt{<PRE>};\pprefix;\texttt{<SUF>};\psp]$;
        \Else
            \State \Comment{Prepare the final output $y$.}
            \State $y \gets [\pprefix;\pmiddle;\psp;\psc]$;
        \EndIf
    \EndFor
    \State {\bfseries Return} output $y$.
    \end{algorithmic}
\end{algorithm}

\section{Experiments}
\label{exp}
In this section, we present extensive experiments to evaluate self-infilling generation across various benchmarks. Please refer to \cref{app:additional_exp_details} for an exhaustive overview of experimental details and \cref{app:additional_exp_results} for additional results including illustrative generation samples.

\subsection{Experimental Setup}
\label{exp:setup}

\paragraph{Benchmarks.} 
Our evaluation encompasses a range of code generation benchmarks, including \humaneval \citep{chen2021codex}, \mbpp \citep{austin2021program}, and \dsonek \citep{lai2023ds1000}. In addition, we also extend our analysis to multilingual code generation with \multiple \citep{cassano2022multiple} and mathematical reasoning with \gsm \citep{cobbe2021gsm}, the detailed results of which can be found at \cref{app:additional_exp_results:gsm8k}.

\paragraph{Code Language Models.} 
For our experiments, we utilize the open-sourced \starcoder \citep{li2023starcoder} and \codellama \citep{roziere2023codellama} models, which have been pre-trained with the FIM objective \citep{bavarian2022fim}. Further model details are available in \cref{app:additional_exp_details:model}.

\paragraph{Evaluation Protocols.}
Following \citet{chen2021codex}, we evaluate the performance of code language models with the pass@$k$ rate, which estimates the probability of a code model generating a correct solution within $k$ attempts. To facilitate a fair comparison to previous work \citep{lai2023ds1000,li2023starcoder,roziere2023codellama}, we report pass@1, pass@10, and pass@100 for the \humaneval and \mbpp benchmarks. We measure pass@1 for other tasks. Pass@1 rates are calculated via greedy decoding, while pass@10 and pass@100 are computed by generating 200 samples at temperature 0.8 using nucleus sampling \citep{Holtzman2020nucleus} with $\text{top-}p = 0.95$. We set the maximum context length to 2048 tokens and limit the maximum number of generated tokens to 512, except for the \humaneval benchmark, where we limit the context length to 640 for accelerating decoding. For self-infilling generation, $\tau$ and $N$ are defaulted to $0.25$ and $2$, respectively, unless otherwise specified.

\begin{table*}[t]
\caption{The pass@1(\%), pass@10(\%), and pass@100(\%) rates on \humaneval (zero-shot) and \mbpp (three-shot) with different code language models. $N$ denotes the number of times the decoding process goes through the loop, and $N\!=\!0$ represents that the looping mechanism is not activated.\textsuperscript{\textdagger} Results are taken from \citet{roziere2023codellama} and \citet{xu2024lemur}.}
\label{tb:humaneval}
\vskip 0.1in
\begin{center}
\resizebox{0.98\linewidth}{!}{ 
\begin{tabular}{lrl|ccc|ccc}
\toprule
\multirow{2}{*}{Model} & \multirow{2}{*}{Size} & \multirow{2}{*}{Method} & \multicolumn{3}{c|}{\humaneval} & \multicolumn{3}{c}{\mbpp}\\
 & & & pass@1 & pass@10 & pass@100 & pass@1 & pass@10 & pass@100\\
\midrule
\multirow{3}{*}{\codellamainst\textsuperscript{\textdagger}} 
& 7B & \multirow{9.8}{*}{Left-to-right} & 34.8 & 64.3 & 88.1 & 44.4 & 65.4 & 76.8 \\
& 13B & & 42.7 & 71.6 & 91.6 & 49.4 & 71.2 & 84.1 \\
& 34B & & 41.5 & 77.2 & 93.5 & 57.0 & 74.6 & 85.4 \\
\cmidrule{1-2} \cmidrule{4-9}
\multirow{3}{*}{\codellamapython\textsuperscript{\textdagger}} & 7B & & 38.4 & 70.3 & 90.6 & 47.6 & 70.3 & 84.8 \\
& 13B & & 43.3 & 77.4 & 94.1 & 49.0 & 74.0 & 87.6 \\
& 34B & & 53.7 & 82.8 & 94.7 & 56.2 & 76.4 & 88.2 \\
\cmidrule{1-2} \cmidrule{4-9}
Lemur\textsuperscript{\textdagger} & 70B & & 35.4  & - & - & 53.2 & - & - \\
GPT-3.5 Turbo\textsuperscript{\textdagger} & - & & 72.6 & - & - & 70.8 & - & - \\
GPT-4 Turbo\textsuperscript{\textdagger} & - & & 88.4 & - & - & 81.0 & - & - \\
\midrule
\midrule %
\multirow{4.5}{*}{\codellama} & \multirow{4.5}{*}{7B}
& Left-to-right & 34.1 & 59.6 & 86.5 & 42.8 & 66.8 & 82.3 \\
\cmidrule{3-9}
& & Self-infill\,($N\!=\!0$) & 29.9 & 56.9 & 86.0 & 41.0 & 67.0 & 84.2 \\
& & Self-infill\,($N\!=\!1$) & 34.1 & 61.3 & 87.0 & 43.8 & 67.6 & 83.9 \\
& & Self-infill\,($N\!=\!2$) & \textbf{39.0} & \textbf{62.5} & \textbf{88.5} & \textbf{44.8} & \textbf{67.9} & \textbf{84.5} \\
\midrule %
\multirow{4.5}{*}{\codellama} & \multirow{4.5}{*}{13B}
& Left-to-right & 35.4 & 69.7 & 90.0 & 47.2 & 71.9 & 87.3 \\
\cmidrule{3-9}
& & Self-infill\,($N\!=\!0$) & 32.3 & 69.2 & 90.5 & 44.0 & 71.2 & 87.8 \\
& & Self-infill\,($N\!=\!1$) & 38.4 & 70.7 & \textbf{92.5} & 47.2 & 72.6 & 88.3 \\
& & Self-infill\,($N\!=\!2$) & \textbf{40.8} & \textbf{72.1} & 91.1 & \textbf{49.0} & \textbf{73.0} & \textbf{89.2} \\
\midrule %
\multirow{4.5}{*}{\starcoderbase} & \multirow{4.5}{*}{15.5B}
& Left-to-right & 31.7 & 56.3 & 80.3 & 43.8 & 68.7 & 85.3 \\
\cmidrule{3-9}
& & Self-infill\,($N\!=\!0$) & 27.4 & 52.0 & 80.5 & 42.2 & 68.7 & 85.7 \\
& & Self-infill\,($N\!=\!1$) & 33.5 & 56.6 & 82.3 & 44.6 & \textbf{69.4} & \textbf{86.4} \\
& & Self-infill\,($N\!=\!2$) & \textbf{36.0} & \textbf{59.4} & \textbf{84.6} & \textbf{46.6} & 69.1 & 84.9 \\
\midrule %
\multirow{4.5}{*}{\starcoder} & \multirow{4.5}{*}{15.5B}
& Left-to-right & 35.4 & 62.1 & 85.1 & 48.6 & \textbf{71.5} & 86.6 \\
\cmidrule{3-9}
& & Self-infill\,($N\!=\!0$) & 29.2 & 58.0 & 85.9 & 46.4 & 70.7 & 85.6 \\
& & Self-infill\,($N\!=\!1$) & 37.8 & 63.2 & 86.5 & 47.8 & 71.3 & 86.1 \\
& & Self-infill\,($N\!=\!2$) & \textbf{38.4} & \textbf{64.7} & \textbf{87.3} & \textbf{50.0} & 71.1 & \textbf{87.2} \\
\bottomrule
\end{tabular}}  
\end{center}
\end{table*}

\subsection{Results}
\label{exp:results}

\paragraph{Results on \humaneval and \mbpp.}
\cref{tb:humaneval} displays the comparative results of self-infilling and traditional decoding approaches on \humaneval and \mbpp benchmarks. When self-infilling is solely equipped with its interruption functionality, without the looping mechanism (denoted as $N=0$), its performance is marginally inferior to vanilla left-to-right completion baselines. This could be attributed to the inherent complexity of infilling tasks compared to next-token prediction, often resulting in difficulties in integrating the prefix with the suffix \citep{bavarian2022fim}. Self-infilling further exacerbates these challenges by requiring the model to join the self-generated suffix with greater variety. 
Nonetheless, as indicated in \cref{exp:analysis}, self-infilling contributes significantly to enhancing the structure of generated code and mitigating potential degenerate issues. Besides, when integrated with the looping mechanism, self-infilling exhibits much higher code generation quality, scales effectively with increased iterations $N$, and consistently outperforms left-to-right baselines. The improvements even sometimes surpass those brought by specialized language training (e.g., \codellamapython) or instruction tuning (e.g., \codellamainst). These results suggest that the performance of \emph{infill-capable} code models can be largely improved by integrating self-infilling capabilities into the inference phase, moving beyond the conventional left-to-right generation paradigm.

\begin{table*}[t]
\caption{Zero-shot pass@1(\%) performance on \dsonek with different code language models. \textsuperscript{\textdagger} Results are taken from \citet{li2023starcoder} and \citet{luo2023wizardcoder}.}
\label{tb:ds1000}
\vskip 0.1in
\centering
\resizebox{0.98\linewidth}{!}{
\begin{tabular}{llcccccccc}
\toprule
\textbf{Model} & \textbf{Method} & \textbf{\rotatebox{35}{\parbox{1.2cm}{\small Matplotlib}}} & \textbf{\rotatebox{35}{\parbox{1.2cm}{\small NumPy}}} & \textbf{\rotatebox{35}{\parbox{1.2cm}{\small Pandas}}} & \textbf{\rotatebox{35}{\parbox{1.2cm}{\small PyTorch}}} & \textbf{\rotatebox{35}{\parbox{1.2cm}{\small SciPy}}} & \textbf{\rotatebox{35}{\parbox{1.2cm}{\small Scikit-Learn}}} & \textbf{\rotatebox{35}{\parbox{1.2cm}{\small TensorFlow}}} & \textbf{Overall} \\
\midrule
CodeGen-16B-Mono\textsuperscript{\textdagger} & Left-to-right & 31.7 & 10.9 & 3.4 & 7.0 & 9.0 & 10.8 & 15.2 & 11.7 \\ 
\midrule
code-cushman-001\textsuperscript{\textdagger} & Left-to-right & 40.7 & 21.8 & 7.9 & 12.4 & 11.3 & 18.0 & 12.2 & 18.1 \\ 
\midrule
code-davinci-001\textsuperscript{\textdagger} & Left-to-right & 41.8 & 26.6 & 9.4 & 9.7 & 15.0 & 18.5 & 17.2 & 20.2 \\
\midrule
\multirow{2}{*}{InCoder-6B\textsuperscript{\textdagger}} & Left-to-right & 28.3 & 4.4 & 3.1 & 4.4 & 2.8 & 2.8 & 3.8 & 7.4 \\
& Insertion & 28.3 & 4.6 & 2.9 & 4.4 & 2.8 & 3.1 & 7.8 & 7.5 \\
\midrule
\multirow{2}{*}{WizardCoder\textsuperscript{\textdagger}} & Left-to-right & 55.2 & 33.6 & 16.7 & 26.2 & 24.2 & 24.9 & 26.7 & 29.2 \\
& Insertion & 55.2 & 35.1 & 20.4 & 30.4 & 28.9 & 32.3 & 37.8 & 32.8 \\
\midrule
\multirow{2}{*}{code-davinci-002\textsuperscript{\textdagger}} & Left-to-right & 57.0 & 43.1 & 26.5 & 41.8 & 31.8 & 44.8 & 39.3 & 39.2 \\ 
& Insertion & 57.0 & 46.5 & 30.1 & 47.7 & 34.8 & 53.7 & 53.4 & 43.3 \\ 
\midrule
\midrule
\multirow{3}{*}{\codellama~7B} 
& Left-to-right & 47.1 & 27.3 & 14.4 & 23.5 & 19.8 & 23.5 & 20.0 & 24.8 \\
& Self-infilling & \textbf{48.4} & 30.0 & \textbf{17.2} & \textbf{27.9} & \textbf{22.6} & \textbf{38.3} & 20.0 & \textbf{28.7} \\ 
& Insertion & 47.1 & \textbf{30.9} & 12.0 & 23.5 & \textbf{22.6} & 33.9 & \textbf{35.6} & 27.1 \\
\midrule
\multirow{3}{*}{\codellama~13B} 
& Left-to-right & 49.0 & 33.2 & 18.6 & 30.9 & \textbf{20.8} & 37.4 & 31.1 & 30.3 \\
& Self-infilling & 49.7 & 33.2 & \textbf{24.4} & 32.4 & \textbf{20.8} & 47.0 & 26.7 & 33.1 \\ 
& Insertion & \textbf{50.3} & \textbf{36.4} & 21.6 & \textbf{33.8} & \textbf{20.8} & \textbf{53.9} & \textbf{35.6} & \textbf{34.4} \\
\midrule
\multirow{3}{*}{\starcoderbase} 
& Left-to-right & \textbf{47.1} & \textbf{31.4} & 9.6 & 26.5 & \textbf{27.4} & 38.3 & 17.8 & 26.9  \\ 
& Self-infilling & 45.8 & 29.1 & 10.0 & 23.5 & 25.5 & \textbf{44.3} & 20.0 & 26.7 \\
& Insertion & 45.8 & \textbf{31.4} & \textbf{12.0} & \textbf{27.9} & 26.4 & 42.6 & \textbf{26.7} & \textbf{28.3} \\ 
\midrule
\multirow{3}{*}{\starcoder} 
& Left-to-right & \textbf{51.6} & 35.0 & 11.3 & 27.9 & 23.6 & 46.1 & 24.4 & 29.8 \\
& Self-infilling & 50.3 & 34.1 & 12.0 & 29.4 & 23.6 & \textbf{47.0} & 26.7 & 29.9 \\ 
& Insertion & 50.3 & \textbf{37.7} & \textbf{13.7} & \textbf{35.3} & \textbf{24.5} & 43.5 & \textbf{31.1} & \textbf{31.5} \\
\bottomrule
\end{tabular}
}
\end{table*}

\paragraph{Results on \dsonek.}
The \dsonek task supports code language models to generate in both \emph{left-to-right} completion and \emph{insertion} formats. In insertion mode, official suffixes are provided for each task (except for Matplotlib problems) to condition generation; while in left-to-right completion mode, these suffixes are translated into succinct natural language specifications and appended to problem descriptions to align with the causal formulation. In our evaluation, we follow the left-to-right format and use the given specifications to construct instance-wise suffix prompts for self-infilling. As shown in \cref{tb:ds1000}, our framework improves the performance over left-to-right completion and narrows the quality gap with the insertion mode even though without the use of official suffixes. Generation examples demonstrate that self-infilling more effectively adheres to the given specifications, such as allocating the final result to a specific variable for evaluation. However, we note that performance gains for \starcoder models are marginal, possibly due to their limited infilling training compared to \codellama series \citep{li2023starcoder,roziere2023codellama}.

\paragraph{Results on Multilingual Code Generation.}
We further evaluate the multilingual performance of self-infilling generation across various programming languages. We recruit the \multiple benchmark \citep{cassano2022multiple} and evaluate our approach for C++, Java, and PHP languages. Detailed results are presented in \cref{app:tb:multiple}, we observe a similar trend in improving conventional left-to-right generation approaches under the multilingual setting. These results indicate the versatility of self-infilling generation across various programming languages.

\begin{table}[t]
\caption{Pass@1(\%) rates on C++, Java, and PHP versions of \humaneval problems from the \multiple benchmark (zero-shot). $N$ denotes the number of times the decoding process goes through the loop, and $N\!=\!0$ represents that the looping mechanism is not activated.}
\label{app:tb:multiple}
\vskip 0.1in
\begin{center}
\resizebox{0.98\linewidth}{!}{
\begin{tabular}{lrl|ccc}
\toprule
\multirow{2}{*}{Model} & \multirow{2}{*}{Size} & \multirow{2}{*}{Method} & \multicolumn{3}{c}{Language} \\
 & & & C++ & Java & PHP \\
 \midrule %
\multirow{4.5}{*}{\starcoderbase} & \multirow{4.5}{*}{15.5B}
& Left-to-right & 30.4 & 27.8 & 25.5 \\
\cmidrule{3-6}
& & Self-infill\,($N\!=\!0$) & 31.1 & 26.6 & 26.1 \\
& & Self-infill\,($N\!=\!1$) & \textbf{31.7} & \textbf{30.4} & 26.1 \\
& & Self-infill\,($N\!=\!2$) & 30.4 & \textbf{30.4} & \textbf{31.7} \\
\midrule %
\multirow{4.5}{*}{\starcoder} & \multirow{4.5}{*}{15.5B}
& Left-to-right & 31.1 & 27.8 & 24.8 \\
\cmidrule{3-6}
& & Self-infill\,($N\!=\!0$) & 31.7 & 27.2 & 25.5 \\
& & Self-infill\,($N\!=\!1$) & 31.1 & \textbf{31.0} & 26.1 \\
& & Self-infill\,($N\!=\!2$) & \textbf{33.5} & 29.7 & \textbf{28.0} \\
\midrule %
\multirow{4.5}{*}{\codellama} & \multirow{4.5}{*}{7B}
& Left-to-right & 28.6 & \textbf{33.5} & 24.2 \\
\cmidrule{3-6}
& & Self-infill\,($N\!=\!0$) & 27.3 & 25.3 & 27.3 \\
& & Self-infill\,($N\!=\!1$) & \textbf{31.7} & 30.4 & \textbf{29.8} \\
& & Self-infill\,($N\!=\!2$) & \textbf{31.7} & 30.4 & 28.0 \\
\midrule %
\multirow{4.5}{*}{\codellama} & \multirow{4.5}{*}{13B}
& Left-to-right & 38.5 & 34.8 & 35.4 \\
\cmidrule{3-6}
& & Self-infill\,($N\!=\!0$) & 36.0 & 34.2 & 29.2 \\
& & Self-infill\,($N\!=\!1$) & 39.1 & 35.4 & \textbf{36.6} \\
& & Self-infill\,($N\!=\!2$) & \textbf{41.0} & \textbf{36.7} & 35.4 \\
\bottomrule
\end{tabular}}  
\end{center}
\end{table}

\subsection{Analysis}
\label{exp:analysis}
\begin{figure}[t]
\centering
\includegraphics[width=\columnwidth]{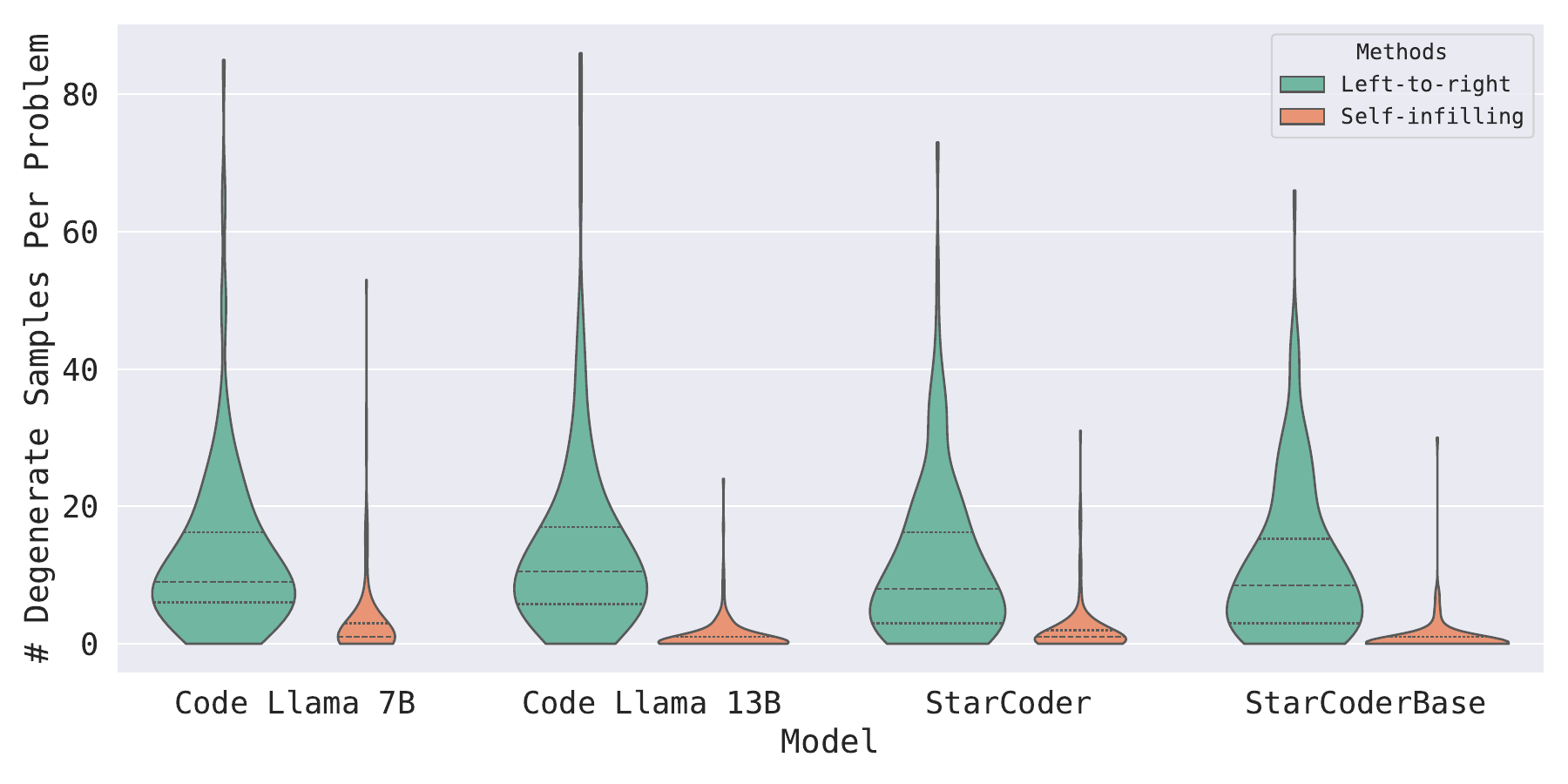}
\caption{The distribution of degenerate solutions from self-infilling ($N\!=\!2$) versus vanilla decoding on \humaneval across various models. For each problem, 200 samples are generated using nucleus sampling with the temperature 0.8 and top-$p$ 0.95.}
\label{fig:degenerate}
\end{figure}

\paragraph{Decoding Regularization.}
As discussed in \cref{model:self_infilling:interruption}, self-infilling facilitates decoding that respects specific constraints, particularly through interruption and suffix prompts. A significant advantage of this approach is its capacity to shape the generation and mitigate degeneracy, a prevalent issue where language models are prone to generating empty or repetitive programs \citep{Holtzman2020nucleus,zhang2023coder-ranker}. We illustrate the effectiveness of such regularization in \cref{fig:degenerate}, which depicts the frequency of \emph{degenerate} samples for both vanilla and self-infilling decoding on \humaneval. Notably, self-infilling significantly reduces the occurrence of degenerate outputs, while vanilla decoding produces more degenerate cases and displays a substantially heavier tail. This improvement is largely attributed to the interruption, which enforces the generation to end with a fitting suffix, such as a return statement in function-level generation. Thanks to the suffix-first non-monotonic formulation, self-infilling effectively shapes the generation toward complete function programs. Our proposed approach makes it easy to implement such regularization while achieving higher performance, which is otherwise difficult to accomplish in conventional left-to-right decoding.

\begin{figure}[t]
\centering
\includegraphics[width=\columnwidth]{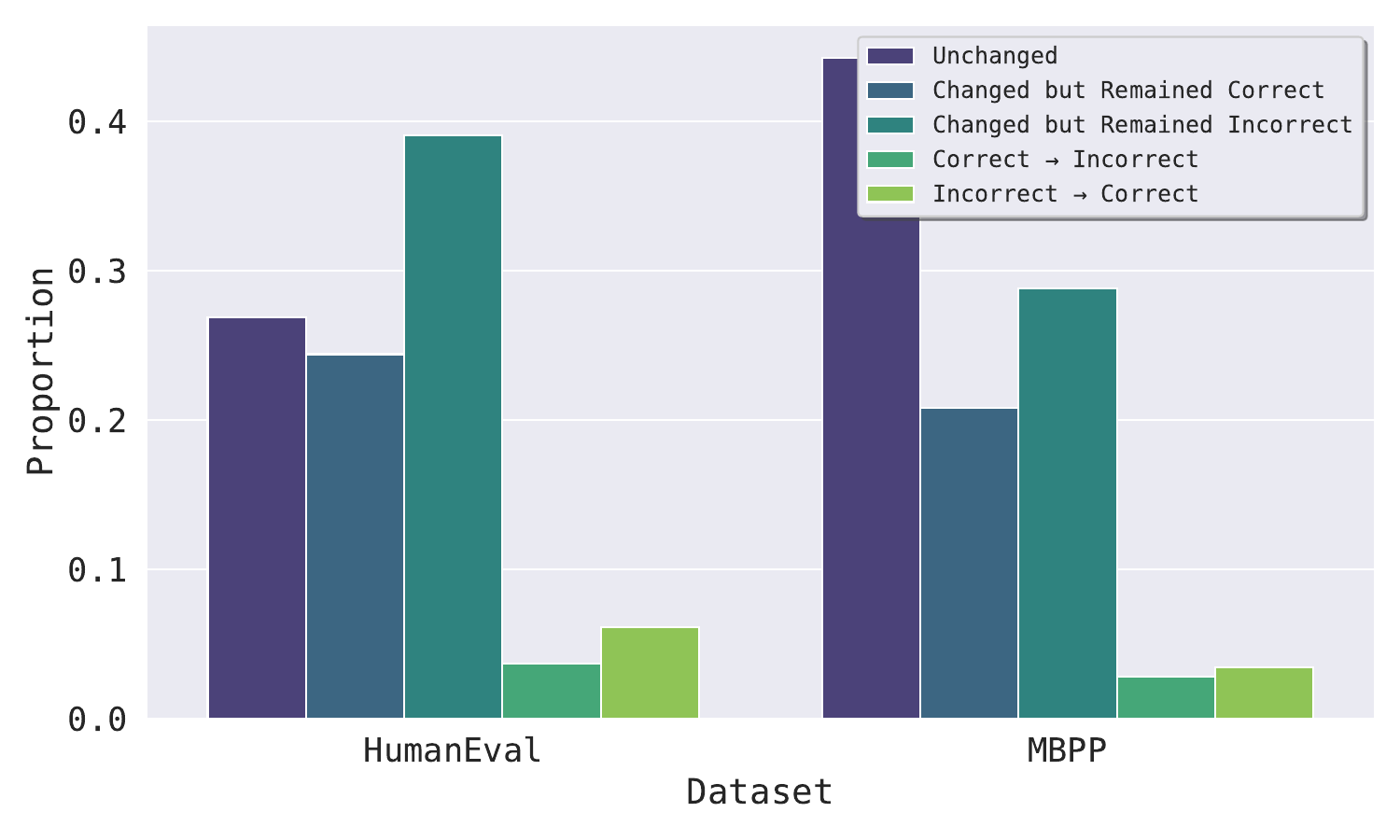}
\caption{Proportional distribution of changes after a second iteration of the looping mechanism ($N\!=\!2$) on \humaneval and \mbpp benchmarks with \codellama~13B. Categories illustrate the state changes of generated code: `\emph{Unchanged}' denotes no change during the second time of looping, `\emph{Changed but Remained Correct/Incorrect}' for changed snippets that stayed correct/incorrect, and `\emph{Correct $\rightarrow$ Incorrect}' for snippets that changed from being correct to incorrect (vice versa).}
\label{fig:loop_inspect}
\end{figure}

\paragraph{Inspecting the Looping Mechanism.}
Different from research on self-improving frameworks \citep{madaan2023selfrefine,chen2023selfdebug,huang2023large} in large language models, our looping mechanism (\cref{model:self_infilling:looped}) operates independently of external tools or self-generated verbal feedback. Instead, the model dynamically modifies the generation \textit{in-place} based on the most recent context. An in-depth examination of this mechanism is provided in \cref{fig:loop_inspect}, which reveals that while overall task performance tends to improve with increasing iterations $N$, the looping mechanism does not intrinsically \emph{improve} code quality. Rather, it simply updates code snippets with more informative contexts and broadens the decoding space, thereby increasing the likelihood of deriving correct solutions on average. It is possible that executing the loop can inadvertently introduce new bugs to initially correct solutions (e.g., the category `\emph{Correct $\rightarrow$ Incorrect}') or fail to fix buggy programs (the `\emph{Changed but Remained Incorrect}' category). We provide additional analyses in \cref{app:additional_exp_results:comp_to_multi_sample} and \cref{app:additional_exp_results:examples}, including illustrative generated examples across various categories (\Cref{fig:example:loop:correct_to_wrong,,fig:example:loop:remain_incorrect,,fig:example:loop:remain_correct,,fig:example:loop:unchanged,,fig:example:loop:wrong_to_correct}).

\subsection{Ablation Study}
\label{exp:ablation}
Additional ablation studies are deferred to \cref{app:additional_exp_results}, including inspecting the effect of varying suffix prompts (\cref{app:additional_exp_results:different_sp}), examining the impact of removing self-infilling from the looping mechanism (\cref{app:additional_exp_results:loop_wo_si}), and comparing looping with sample-and-rank approaches (\cref{app:additional_exp_results:comp_to_multi_sample}).

\begin{figure}[t]
\centering
\includegraphics[width=\columnwidth]{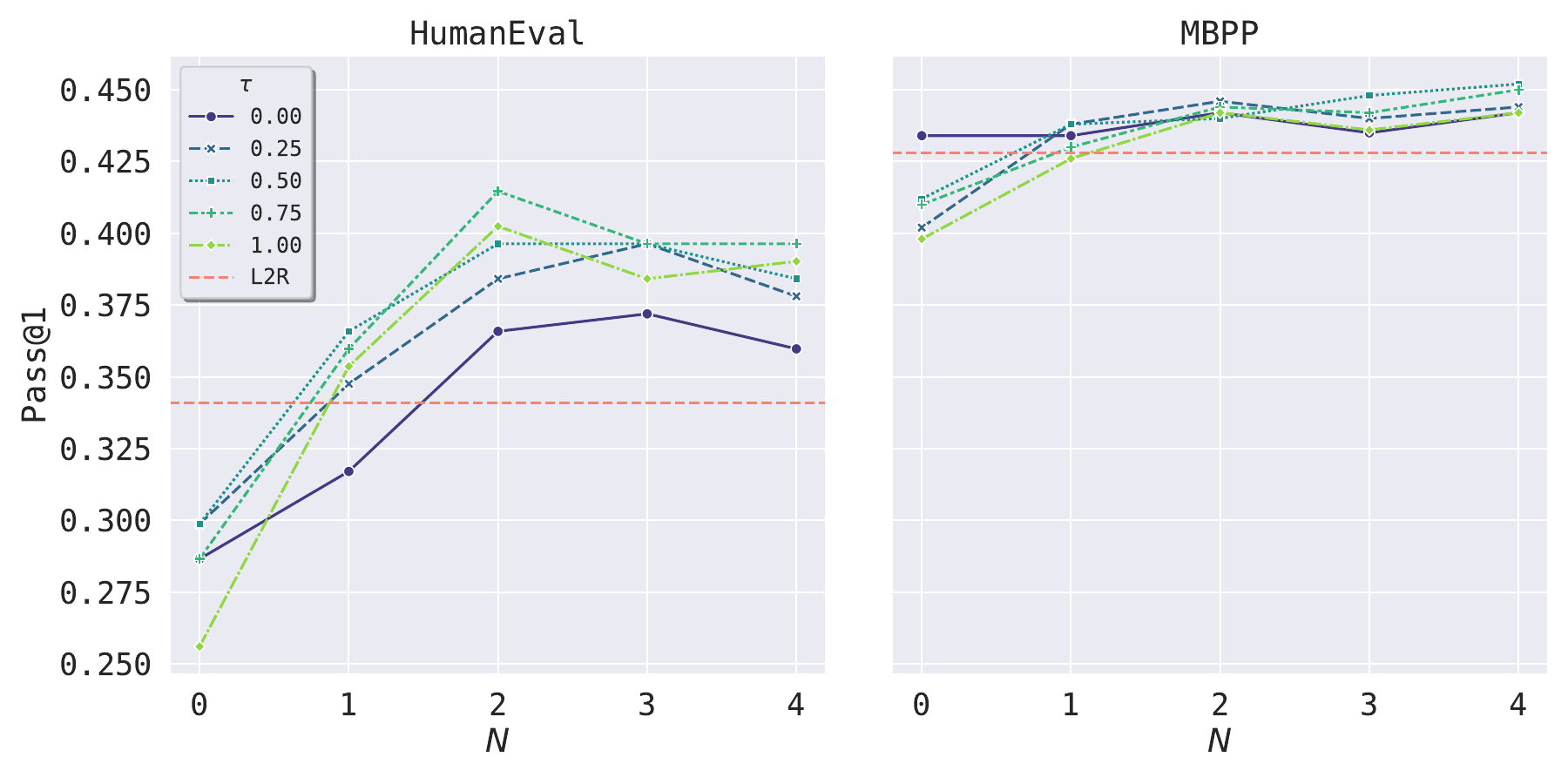}
\caption{Results on \humaneval and \mbpp with different values of $\tau$ and $N$ on \codellama~7B. $N=0$ indicates that the looping mechanism is disabled, and the horizontal dashed line represents the performance of the vanilla left-to-right baseline (L2R).}
\label{fig:ablation:tau}
\end{figure}

\paragraph{On the Effect of $\tau$ and $N$.}
$\tau$ and $N$ are main hyper-parameters in our self-infilling generation framework. The parameter $\tau$ determines the trigger point for the self-infilling interruption. Specifically, a larger $\tau$ value indicates a more aggressive approach towards activating self-infilling to regularize generation, and vice versa. Besides, the parameter $N$ controls the duration of looped decoding execution. We conduct a detailed analysis to examine the influence of $\tau$ and $N$ on the generation results, as presented in \cref{fig:ablation:tau} (as well as \cref{fig:ablation:tau_n_codellama13b,,fig:ablation:tau_n_starcoder,,fig:ablation:tau_n_starcoderbase} in \cref{app:additional_exp_results:ablation_tau_n}).
When the looping mechanism (\cref{model:self_infilling:looped}) is inactive ($N=0$), We observe that a larger $\tau$ typically correlates with slightly worse performance. This could be attributed to \textbf{1)} the complexity of infilling and \textbf{2)} the challenge in crafting an apt suffix according to a limited prefix; nevertheless, large $\tau$ values lead to better control over generation structures. For instance, nearly 13\% of solutions generated by \codellama~7B on \humaneval exhibit degeneracy at $\tau = 0.1$, which is reduced to 2.4\% when increasing $\tau$ to $0.25$. When equipped with the looping mechanism, most $\tau$ settings exhibit improved performance as $N$ increases, despite some fluctuations at longer looping durations. Their performance generally surpasses the left-to-right completion baseline significantly. An exception is observed when $\tau = 0.0$, where self-infilling is not engaged in the first iteration. This case is equivalent to initializing the generation with left-to-right decoding, which is shown to benefit less from looping compared to other settings of $\tau$. This trend may step from the inherently less structured nature of the left-to-right generation, narrowing the exploration space in subsequent iterations.

\paragraph{On the Implementation of Looping Mechanisms.}
The introduced looping mechanism iteratively updates snippets of output to enhance decoding. A pivotal aspect of these iterations involves the transmission of updated contexts to the subsequent iteration (\cref{alg:loop}). Unlike self-infilling, our mechanism's left-to-right decoding lacks a clear notion of \psp and \psc, necessitating strategic output parsing to extract essential pieces for the next update cycle. We explore three distinct strategies for extraction: \textbf{1)} \textbf{Vanilla}, where the new \psp starts at the beginning of updated \psuffix and ends at the occurrence of the default \psp; \textbf{2)} \textbf{Extended}, similar to Vanilla but starting from the midpoint of the whole generation to enlarge \psp; and \textbf{3)} \textbf{Half}, splitting the entire generation uniformly and employing the latter half as \psp, where the self-infilling reduces to standard infilling for \cref{eqn:si:middle}. These strategies reflect an increased amount of context information available for the subsequent iteration.

Our analysis (\cref{tb:ablation:suffix_split}) indicates that \codellama models usually exhibit greater resilience to splitting strategy variations. Conversely, models like \starcoder generally benefit from a more informative suffix prompt, encompassing a broader context. This might step from less sufficient infilling training in \starcoder (only 50\% of its pre-training time) compared to that in \codellama (90\% of pre-training), which restricts its capability of (self-)infilling according to a limited context. In addition, the Half strategy, despite being more informative, often leads to slightly inferior results. This could be due to its complete suffix being overly specific and narrowing the solution space, compared to the other strategies that allow for more model-drive completion. Through our experiments, we utilize the Extended strategy for all code language models, except for \codellama~7B where strategy Vanilla is used.

\begin{table}[t]
\caption{Results on \humaneval and \mbpp with different suffix splitting strategies in the looping mechanism.}
\label{tb:ablation:suffix_split}
\begin{center}
\resizebox{0.98\columnwidth}{!}{ 
\begin{tabular}[t]{ll|cccc}
\toprule
\multirow{3}{*}{Dataset} & \multirow{3}{*}{Strategy} & \multicolumn{4}{c}{Model} \\
& & \multicolumn{2}{c}{\starcoder} & \multicolumn{2}{c}{\codellama} \\
& & 15B-\textsc{Base} & 15B & 7B & 13B \\
\midrule
\multirow{3}{*}{\humaneval}
& Vanilla  & 31.7          & 34.8 &  \textbf{37.2} & 35.9 \\
& Extended & 36.0          & 37.8 & 32.3 & \textbf{40.8}  \\
& Half     & \textbf{37.8} & \textbf{38.4}  & 31.7 & 39.6 \\
\midrule
\multirow{3}{*}{\mbpp}
& Vanilla  & 41.6 & 44.8 & \textbf{44.6} & \textbf{49.0} \\
& Extended & \textbf{46.6} & \textbf{50.8} & 43.8 & \textbf{49.0}  \\
& Half     & 43.8 & 48.8  & 43.8 & 46.8  \\
\bottomrule
\end{tabular}} 
\end{center}
\end{table}

\section{Related Work}
\label{related_work}
A variety of language models possess the capability to perform infilling \citep{sun2024survey}. Encoder-only \citep{devlin2018bert,liu2019roberta,joshi2020spanbert} and encoder-decoder architectures \citep{raffel2020t5,lewis-etal-2020-bart,aghajanyan2021htlm,tay2023ul2} are capable of conditioning on the bidirectional context, but they primarily focus on representation learning and fall short of generating coherent content. Standard left-to-right causal language models excel in generating high-quality text \citep{radford2018gpt,radford2019gpt2,brown2020gpt3}; however, their inherent causal formulation limits their effectiveness in infilling tasks that require bidirectional context. To circumvent these limitations, several approaches tailor the auto-regressive model architecture to generate tokens in a more flexible order beyond the standard left-to-right direction \citep{yang2019xlnet}. Techniques such as integrating both left-to-right and right-to-left language models \citep{nguyen2023mim} have been effective in capturing bidirectional dependencies and thus facilitating infilling. Additionally, the ordering of generation can be made adaptive based on the model output or enhanced by learning a location predictor for each token \citep{stern2019insertion,gu2019indigo,chan2019kermit,welleck2019non,shen2020blank,alon2020structural,shen2023film}, further enhancing model flexibility in generation.

Another line of research enables infilling by transforming the input sequences while \emph{retaining} the left-to-right auto-regressive formulation \citep{zhu2019textinfilling,donahue-etal-2020-enabling,tay2022upalm,du2022glm,aghajanyan2022cm3,bavarian2022fim,fried2023incoder}. These methods reformat the input sequence by randomly selecting various spans and relocating them to the end of the sequence. The language model is then trained to predict tokens in the standard left-to-right auto-regressive manner but under this permuted sequence, which learns to infill considering both preceding and following content. This conceptually simple framework can be considered as extending span corruption objectives \citep{raffel2020t5,tay2023ul2,tay2022upalm,anil2023palm2}, which are commonly used in training encoder-decoder Transformers, to the context of decoder-only language models. For instance, the causal masking objective \citep{aghajanyan2022cm3,yasunaga2023racm3,fried2023incoder,yu2023cm3leon,nijkamp2023codegen2} samples a number of contiguous token spans at random, moves these spans to the end of the input sequence, and replaces the tokens at the original position with mask sentinel tokens. GLM \citep{du2022glm,zeng2022glm130b} further generalizes the span corruption objective by randomly permuting the order among different spans to fully capture the inter-dependencies between different spans. Fill-In-the-Middle \citep[FIM; ][]{bavarian2022fim} employs a similar form as the causal masking objective but only samples a single span. These objectives are specially recruited in training several code language models, including recent versions of Codex \citep{edit_insert}, \textsc{InCoder} \citep{fried2023incoder}, \textsc{SantaCoder} \citep{allal2023santacoder}, \starcoder \citep{li2023starcoder}, \starcoder~2 \citep{lozhkov2024starcoder2}, \textsc{CodeGen~2/2.5} \citep{nijkamp2023codegen2}, \codellama \citep{roziere2023codellama}, DeepSeek-Coder \citep{guo2024deepseekcoder}, and CodeGemma \citep{codegemma2024}, facilitating numerous downstream tasks including partial code completion, docstring generation, return type prediction, and adaptive retrieval-augmented generation \citep{wu2024repoformer}.

In this work, we extend the study of FIM objectives used in training language models and investigate their \emph{built-in} self-infilling capability. Complementary to prior efforts in infilling training objectives, our work explores the advantages of imbuing infilling capabilities with the \emph{decoding} process. While previous findings indicate that infilling can be learned in pre-training without (or slightly) compromising left-to-right generation quality \citep{bavarian2022fim,li2023starcoder,nijkamp2023codegen2,roziere2023codellama}, our findings suggest decoding can be much enhanced by incorporating the acquired infilling capability.

\section{Conclusion}
\label{conclusion}
This work explores the built-in self-infilling capability of FIM-trained code language models, based on which we develop a code generation framework that integrates infilling with auto-regressive decoding. Our method extends traditional decoding to a non-monotonic process that supports interruption and looping mechanisms, allowing the model to defer the generation of some contexts and recursively update code snippets cyclically. Throughout extensive experiments, we demonstrate that self-infilling decoding significantly improves generation quality and regularity.

\paragraph{Limitations and Future Directions.}
Our findings suggest that the decoding behavior of language models can be effectively programmed and extended by harnessing their (self-)infilling abilities. This highlights the considerable potential of language models trained with diverse objectives like FIM \citep{bavarian2022fim}, which not only maintain scalability \citep{bavarian2022fim,tay2022upalm,tay2023ul2,anil2023palm2} but also yield possibly improved generation quality. Besides, there are several interesting directions for future work, some of which we outline below:

\begin{itemizesquish}
    \item While our framework is primarily tailored for code generation tasks, its application to other domains, such as mathematical reasoning, offers an intriguing avenue for future exploration.
    \item The developed interruption and looping techniques present our initial attempts to exploit self-infilling. There exists potential to guide language models towards more structured generation, such as conforming to context-free grammars \citep{willard2023outlines,microsoft2023guidance}.
    \item Our self-infilling framework is limited to single-span infilling due to the formulation of FIM objectives. Extending self-infilling to accommodate multiple or nested spans is a compelling direction for future research.
    \item The looping mechanism developed in this work incurs additional computational overhead due to repeated context processing and decoding operations. Future work might include optimization of key-value caching and reuse across iterations to enhance efficiency.
\end{itemizesquish}

\section*{Acknowledgements}
We would like to thank the HKU NLP group and the anonymous reviewers for their valuable suggestions that greatly helped improve this work. Our manuscript especially benefited from insightful
discussions with Yiheng Xu, Tianbao Xie, Chenxin An, and Hongjin Su. This work is partially supported by the joint research scheme of the National Natural Science Foundation of China (NSFC) and the Research Grants Council (RGC) under grant number N\_HKU714/21.

\section*{Impact Statement}
This paper presents work whose goal is to advance the field of Machine Learning. There are many potential societal consequences of our work, none of which we feel must be specifically highlighted here.

\bibliography{paper}
\bibliographystyle{icml2024}

\newpage
\appendix
\onecolumn
\textbf{\huge Appendices}\vspace{0.05in}

\section{Pseudo-code}
\label{app:pseudocode}
In practice, we implement our self-infilling interruption technique (\cref{model:self_infilling:interruption}; \cref{alg:self-infill}) in a stateless manner through the manipulation of sentinel tokens. For completeness, we also include the pseudo-code of the conventional left-to-right generation process used in our looping mechanism (\cref{model:self_infilling:looped}), as shown in \cref{alg:left-to-right}. Note that in both \cref{alg:self-infill,alg:left-to-right} we omit detailed decoding setups, such as the use of temperature scaling and nucleus sampling.

\begin{algorithm}[ht] %
   \caption{Self-infilling Interruption}
   \label{alg:self-infill}
    \begin{algorithmic}
    \State {\bfseries Input:} Input sequence $x$, language model $f(\cdot;\vtheta)$, suffix prompt tokens $\psp$, stop condition $\sS$, and probability threshold $\tau$.
    \State {\bfseries Output:} Self-infilled output.
    \State
    \For{$t = 1,2,\dots$}
        \State Calculate the probability distribution over the next token $p(x_t) \gets f(x;\vtheta)$;
        \If{\texttt{<PRE>} $\in x$ and \texttt{<SUF>} $\notin x$ and \texttt{<MID>} $\notin x$}
            \State \Comment{Prefix generation.}
            \If{$\max_v p(v) < \tau$}
                \State \Comment{In case the model becomes uncertain, interrupt and pivot to suffix generation.}
                \State Set $p(x_t) = \delta_{x_t}(\texttt{<SUF>})$ to place all probability mass on sentinel token \texttt{<SUF>};
            \EndIf
        \ElsIf{\texttt{<PRE>} $\in x$ and \texttt{<SUF>} $\in x$ and \texttt{<MID>} $\notin x$}
            \State \Comment{Suffix generation.}
            \State Calculate the length of the current suffix $l_\texttt{suffix}$ and the suffix prompt $l_\psp$;
            \If{$l_\texttt{suffix} < l_\psp$}
                \State \Comment{Override the next token logit so that the suffix has to start with $\psp$.}
                \State Set $p(x_t) = \delta_{x_t}(\psp[l_\texttt{suffix}])$;
            \EndIf
        \EndIf
        \State \Comment{Middle generation (i.e., infilling) does not require extra post-processing.}
        \State
        \If{stop condition $\sS$ are met}
            \If{\texttt{<PRE>} $\in x$ and \texttt{<SUF>} $\in x$ and \texttt{<MID>} $\notin x$}
                \State \Comment{If the suffix is being generated, populate \texttt{<MID>} to signify the start of infilling.}
                \State Set $p(x_t) = \delta_{x_t}(\texttt{<MID>})$;
            \Else
                \State \Comment{Otherwise, the generation process is terminated.}
                \State {\bfseries break}
            \EndIf
        \EndIf
        \State Draw the next token $x_t \sim p(x_t)$ and append $x_t$ to the current input $x \gets [x;x_t]$;
    \EndFor
    \State {\bfseries Return} the updated sequence $x$.
    \end{algorithmic}
\end{algorithm}

\begin{algorithm}[ht] %
   \caption{Left-to-right Generation}
   \label{alg:left-to-right}
    \begin{algorithmic}
    \State {\bfseries Input:} Input sequence $x$, language model $f(\cdot;\vtheta)$, and a list of stop tokens $\sS$.
    \State {\bfseries Output:} Left-to-right decoded output.
    \For{$t = 1,2,\dots$}
        \State Calculate the probability distribution over the next token $p(x_t) \gets f(x;\vtheta)$;
        \State
        \If{stop tokens $\sS$ are met}
            \State \Comment{The generation process is terminated if the current generation contains stop tokens.}
            \State {\bfseries break}
        \EndIf
        \State Draw the next token $x_t \sim p(x_t)$ and append $x_t$ to the current input $x \gets [x;x_t]$;
    \EndFor
    \State {\bfseries Return} the updated sequence $x$.
    \end{algorithmic}
\end{algorithm}

\begin{figure}[ht!]
\begin{DSOuterBox}{Parsing Function for Left-to-right Outputs}
\begin{codelisting}{python}
def l2r_parser(x, prompt, suffix_p, strategy):
    # x: The current output sequence
    # prompt: The original input prompt
    # suffix_p: The suffix prompt
    # strategy: The strategy used for splitting the output, as in Table 3 of Section 3.4.

    l_prompt = x.find(prompt) + len(prompt)
    # the index of midpoint of the whole generation excluding the original prompt
    l_half = l_prompt + (len(x) - l_prompt) // 2
    l_suffix_p_start = x.find(suffix_p, l_half)
    l_suffix_p_end = l_suffix_p_start + len(suffix_p)

    prefix = x[:l_prompt]
    if strategy == "Vanilla":
        middle   = x[l_prompt:l_suffix_p_start]
        suffix_p = x[l_suffix_p_start:l_suffix_p_end]
        suffix_c = x[l_suffix_p_end:]
    elif strategy == "Extended":
        middle   = x[l_prompt:l_half]
        suffix_p = x[l_half:l_suffix_p_end]
        suffix_c = x[l_suffix_p_end:]
    elif strategy == "Half":
        middle   = x[l_prompt:l_half]
        suffix_p = x[l_half:]
        suffix_c = ""

    return (prefix, middle, suffix_p, suffix_c)
\end{codelisting}
\end{DSOuterBox}
\caption{Python pseudo-code implementation of the parsing function for the left-to-right generation.}
\label{fig:parse-function}
\end{figure}

\section{Additional Experimental Details}
\label{app:additional_exp_details}
\subsection{Task Details}
\label{app:additional_exp_details:task}
\begin{itemizesquish}
\item \textbf{\humaneval} \citep{chen2021codex} consists of 164 crafted programming challenges, each accompanied by multiple unit tests to evaluate the correctness of solutions generated by code models. We conduct evaluations in a zero-shot manner. Following previous practice \citep{li2023starcoder}, for \starcoder series, we strip off the trailing newline symbol \mintinline{python}{\n} appearing at the end of the official prompt to align with their trained tokenizer \citep{microsoft2023guidance}; we adhere to the official prompt format in evaluation for the remaining models.
\item \textbf{\multiple} \citep{cassano2022multiple} is a multilingual benchmark extending \humaneval to various programming languages. In this work, we report results of self-infilling generation for C++, Java, and PHP languages.
\item \textbf{\mbpp} \citep{austin2021program} includes 500 crowd-sourced basic Python programming problems as the test set to evaluate code models. Following \citet{austin2021program}, we use their provided 3-shot prompts for both \starcoder and \codellama model families.
\item \textbf{\dsonek} \citep{lai2023ds1000} comprises 1,000 data science questions sourced from StackOverflow, aimed at benchmarking code models against real-world scenarios. The questions span various Python libraries commonly used in data science, including Matplotlib (155 questions), NumPy (220), Pandas (291), PyTorch (68), SciPy (106), Scikit-learn (115), and Tensorflow (45). We use the provided prompt to perform zero-shot evaluation. There are two distinct prompt formats in \dsonek available for models considered in this work: the \emph{left-to-right completion} format, which puts the entire instruction in the left context for regular left-to-right decoding; and the \emph{insertion} format, which provides the instruction in both the left and right contexts. Our self-infilling mode follows the left-to-right format for generation, with further details elaborated below.
\item \textbf{\gsm} \citep{cobbe2021gsm} comprises various grade school math word problems. We assess the performance of language models on the \gsm test set, which includes 1,319 instances, using an 8-shot in-context example prompt.
\end{itemizesquish}

\begin{table}[t]
\caption{The pass@1(\%), pass@10(\%), and pass@100(\%) rates on \humaneval (zero-shot) with \deepseekcoder models. $N$ denotes the number of times the decoding process goes through the loop, and $N\!=\!0$ represents that the looping mechanism is not activated.}
\label{app:tb:deepseek:humaneval}
\vskip 0.1in
\begin{center}
\resizebox{0.65\linewidth}{!}{
\begin{tabular}{lrl|ccc}
\toprule
\multirow{2}{*}{Model} & \multirow{2}{*}{Size} & \multirow{2}{*}{Method} & \multicolumn{3}{c}{\humaneval} \\
 & & & pass@1 & pass@10 & pass@100 \\
\midrule %
\multirow{4.5}{*}{\deepseekcoder} & \multirow{4.5}{*}{1.3B}
& Left-to-right & 31.1 & 52.0 & 78.3 \\
\cmidrule{3-6}
& & Self-infill\,($N\!=\!0$) & 27.4 & 48.4 & 72.7 \\
& & Self-infill\,($N\!=\!1$) & 32.3 & 52.9 & 79.1 \\
& & Self-infill\,($N\!=\!2$) & \textbf{35.4} & \textbf{54.6} & \textbf{80.1} \\
\midrule %
\multirow{4.5}{*}{\deepseekcoder} & \multirow{4.5}{*}{6.7B}
& Left-to-right & 47.0 & 75.2 & 92.1 \\
\cmidrule{3-6}
& & Self-infill\,($N\!=\!0$) & 36.6 & 70.5 & 89.6 \\
& & Self-infill\,($N\!=\!1$) & 46.3 & 76.2 & 92.1 \\
& & Self-infill\,($N\!=\!2$) & \textbf{48.2} & \textbf{78.4} & \textbf{92.4} \\
\midrule %
\multirow{4.5}{*}{\deepseekcoder} & \multirow{4.5}{*}{33B}
& Left-to-right & 49.4 & 81.4 & 93.1 \\
\cmidrule{3-6}
& & Self-infill\,($N\!=\!0$) & 49.4 & 78.9 & 92.2 \\
& & Self-infill\,($N\!=\!1$) & \textbf{58.5} & 83.0 & 94.1 \\
& & Self-infill\,($N\!=\!2$) & \textbf{58.5} & \textbf{84.0} & \textbf{94.8} \\
\bottomrule
\end{tabular}}  
\end{center}
\end{table}

\begin{table*}[t]
\caption{Zero-shot pass@1(\%) performance on \dsonek with \deepseekcoder models. \textsuperscript{\textdagger} Results are taken from \citet{guo2024deepseekcoder}.}
\label{app:tb:deepseek:dsonek}
\vskip 0.1in
\centering
\resizebox{0.95\linewidth}{!}{
\begin{tabular}{llcccccccc}
\toprule
\textbf{Model} & \textbf{Method} & \textbf{\rotatebox{35}{\parbox{1.2cm}{\small Matplotlib}}} & \textbf{\rotatebox{35}{\parbox{1.2cm}{\small NumPy}}} & \textbf{\rotatebox{35}{\parbox{1.2cm}{\small Pandas}}} & \textbf{\rotatebox{35}{\parbox{1.2cm}{\small PyTorch}}} & \textbf{\rotatebox{35}{\parbox{1.2cm}{\small SciPy}}} & \textbf{\rotatebox{35}{\parbox{1.2cm}{\small Scikit-Learn}}} & \textbf{\rotatebox{35}{\parbox{1.2cm}{\small TensorFlow}}} & \textbf{Overall} \\
\midrule
\multirow{4}{*}{\deepseekcoder~1.3B}
& Left-to-right\textsuperscript{\textdagger} & 32.3 &	21.4 &	9.3 &	8.8 &	8.5 & 	16.5 & 	8.9	& 16.2 \\
& Left-to-right	& 35.5 &	22.3 &	9.6 &	7.4 &	7.5	 & 27.8	 & 11.1	& 18.2 \\
& Self-infilling & 36.8 &	21.4 &	9.6 &	7.4 &	11.3	 & 27.0	 & 11.1	& 18.5 \\
& Insertion	& 35.5 & 20.9 &	5.2 &	7.4 &	13.2	 & 13.0	 & 15.6	& 15.7 \\
\midrule
\multirow{4}{*}{\deepseekcoder~6.7B}
& Left-to-right\textsuperscript{\textdagger}	& 48.4 & 35.5 & 20.6 & 19.1 & 22.6 & 38.3 & 24.4 & 30.5 \\
& Left-to-right	& 51.6 & 39.1 & 22.3 & 19.1 & 24.5 & 45.2 & 26.7 & 33.4 \\
& Self-infilling	& 52.3 & 38.6 & 26.8 & 29.4 & 23.6 & 42.6 & 28.9 & 35.1 \\
& Insertion	& 51.6 & 45.9 & 31.3 & 22.1 & 29.2 & 42.6 & 40.0 & 38.5 \\
\midrule
\multirow{4}{*}{\deepseekcoder~33B}
& Left-to-right\textsuperscript{\textdagger}	& 56.1 & 49.6 & 25.8 & 36.8 & 36.8 & 40.0 & 46.7 & 40.2 \\
& Left-to-right	& 60.6 & 51.8 & 28.2 & 35.3 & 34.9 & 48.7 & 46.7 & 42.8 \\
& Self-infilling	& 61.9 & 52.3 & 28.9 & 44.1 & 34.0 & 47.8 & 46.7 & 43.7 \\
& Insertion	& 61.3 & 49.5 & 32.0 & 47.1 & 29.2 & 47.0 & 60.0 & 44.1 \\
\bottomrule
\end{tabular}
}
\end{table*}

\subsection{Model Details}
\label{app:additional_exp_details:model}
For code language models, our study mainly utilizes \starcoderbase and \starcoder from the \starcoder family \citep{li2023starcoder}, along with \codellama~7B and \codellama~13B from the \codellama series \citep{roziere2023codellama}, respectively. Note that we do not evaluate \codellama~34B as it is not trained with the infilling objective \citep{roziere2023codellama}. 

\paragraph{Self-infilling with \deepseekcoder Models.}
We also evaluate the performance of \deepseekcoder families \citep{guo2024deepseekcoder}, a series of recently released infilling-capable code language models. As detailed in \cref{app:tb:deepseek:humaneval} and \cref{app:tb:deepseek:dsonek}, we observe significant improvements in performing self-infilling generation compared to conventional left-to-right approaches. Notably, this advantage scales well with the model size, where \deepseekcoder~33B even demonstrates a 9\% absolute increase in pass@1 rate on the \humaneval benchmark. On \dsonek, the results are consistent with the trends observed for \codellama with \deepseekcoder 6.7B and 33B models. In particular, our approach effectively narrows the gap between left-to-right completion and the insertion baseline (which provides more informative bidirectional contexts). However, \deepseekcoder~1.3B exhibits a divergence from this trend, with the insertion baseline yielding slightly worse performance. This difference might be attributed to the model’s smaller size, limiting its ability to generalize infilling capabilities to more diverse problems.

\begin{wraptable}[22]{R}{0.5\columnwidth}
\caption{The pass@1(\%) rates on \humaneval and \humanevalplus (both are evaluated zero-shot) benchmarks with different instruct code models.}
\label{app:tb:supp-main-results}
\vskip 0.1in
\begin{center}
\resizebox{0.97\linewidth}{!}{ 
\begin{tabular}{lrl|cc}
\toprule
Model & Size & Method & \humaneval & \humanevalplus \\
\midrule %
\multirow{4.5}{*}{\codellamainst} & \multirow{4.5}{*}{7B}
& Left-to-right & 43.9 & 38.4 \\
\cmidrule{3-5}
& & Self-infill\,($N\!=\!0$) & 42.7 & 37.2 \\
& & Self-infill\,($N\!=\!1$) & 45.1 & 39.6 \\
& & Self-infill\,($N\!=\!2$) & \textbf{45.7} & \textbf{40.9} \\
\midrule %
\multirow{4.5}{*}{\codellamainst} & \multirow{4.5}{*}{13B}
& Left-to-right & \textbf{43.3} & \textbf{36.6} \\
\cmidrule{3-5}
& & Self-infill\,($N\!=\!0$) & 37.2 & 31.1 \\
& & Self-infill\,($N\!=\!1$) & 42.1 & 34.8 \\
& & Self-infill\,($N\!=\!2$) & 42.1 & 35.4 \\
\midrule %
\multirow{4.5}{*}{\deepseekcoderinst} & \multirow{4.5}{*}{1.3B}
& Left-to-right & \textbf{68.3} & \textbf{63.4} \\
\cmidrule{3-5}
& & Self-infill\,($N\!=\!0$) & 57.3 & 52.4 \\
& & Self-infill\,($N\!=\!1$) & 65.2 & 59.8 \\
& & Self-infill\,($N\!=\!2$) & 65.2 & 58.5 \\
\midrule %
\multirow{4.5}{*}{\deepseekcoderinst} & \multirow{4.5}{*}{6.7B}
& Left-to-right & 78.7 & 70.1 \\
\cmidrule{3-5}
& & Self-infill\,($N\!=\!0$) & 73.8 & 66.5 \\
& & Self-infill\,($N\!=\!1$) & \textbf{79.3} & \textbf{72.0} \\
& & Self-infill\,($N\!=\!2$) & 78.7 & \textbf{72.0} \\
\midrule %
\multirow{4.5}{*}{\deepseekcoderinst} & \multirow{4.5}{*}{33B}
& Left-to-right & 77.4 & 68.9 \\
\cmidrule{3-5}
& & Self-infill\,($N\!=\!0$) & 74.4 & 67.7 \\
& & Self-infill\,($N\!=\!1$) & \textbf{79.3} & \textbf{71.3} \\
& & Self-infill\,($N\!=\!2$) & 77.4 & 69.5 \\
\bottomrule
\end{tabular}}  
\end{center}
\end{wraptable}

\paragraph{Self-infilling with Instruct Models.}
While these model families encompass other variants supporting infilling, such as \codellamainst, it is much more complicated to deal with the interplay between the instruction and infilling sentinel tokens, which might override the effect of one with the other. To examine this, we conduct evaluations of these instruct models, in particular \codellamainst and \deepseekcoderinst, on \humaneval as well as \humanevalplus \citep{evalplus} for a more rigorous and robust evaluation. Besides, we prepare the input prompt according to these models' respective formats. As detailed in \cref{app:tb:supp-main-results}, our findings reveal mixed results when applying self-infill generation to these language models. Generally, self-infilling does not significantly improve the performance of these instruct models, nor does increasing the looping time yield substantial benefits. This can be attributed to the \emph{absence} of infilling training in the instruction fine-tuning stage, complicating the simultaneous application of instruction formatting and infilling patterns during decoding. This observation suggests a potential area for future research: exploring the usage of incorporating FIM during instruction fine-tuning to further improve self-infilling generation.

\paragraph{Self-infilling with More Flexible Infilling-capable Models.}
Notably, there are other open-sourced code models like \textsc{InCoder} \citep{fried2023incoder} and \textsc{CodeGen~2/2.5} \citep{nijkamp2023codegen2} compatible with multi-span infilling. However, our preliminary experiments indicate these models are difficult to perform self-infilling, particularly in generating a coherent suffix from a given prefix prompt. This limitation may stem from their training paradigms, which involve more intricate span corruption objectives \citep{aghajanyan2022cm3,tay2023ul2} as opposed to the FIM objective \citep{bavarian2022fim}. In their input sequence construction, the same set of sentinel tokens marks both the suffix and middle segments. For example, a typical training instance might follow the format \texttt{[prefix; <Mask:0>; suffix; <Mask:0>; middle; <EOT>]}, with \texttt{<Mask:0>} indicating the start of both the suffix and middle segments. Consequently, these models may struggle to discern whether the ongoing generation pertains to the suffix or the infilled section. As a result, our study focuses on the \starcoder and \codellama families, deferring an extensive analysis of more flexible infilling models to future work.

\subsection{Self-infilling Implementation Details}
\label{app:additional_exp_details:impl}
In this section, we provide comprehensive details of our self-infilling implementation.

\paragraph{The Use of Stopping Criteria.}
Common code language models usually necessitate specific stopping conditions to appropriately terminate decoding, since they struggle to faithfully stop their generation as expected. This is often accomplished by monitoring for specific tokens in the generated code, such as incorporating markers like \texttt{<code>} and \texttt{</code>}; the decoding process is then terminated upon encountering the closing marker \texttt{</code>}. Another example is employing stop tokens indicative of the start of a new function, class implementation, or assertion during function-level generation. In our looping mechanism, where multiple rounds of decoding occur, we terminate both self-infilling and left-to-right generation upon these stop tokens and remove any extra content after the first met stop token.

\paragraph{Tokenization.}
Proper tokenization in infill-capable language models often poses challenges \citep{microsoft2023guidance,roziere2023codellama}, especially when different parts of the generation output (\pprefix, \pmiddle, and \psuffix) may break up tokens across piece boundaries. This can sometimes lead to irregular tokens and hurt performance due to out-of-distribution tokens. We apply a heuristic strategy to alleviate this by right-stripping all spaces from the current generation output, making the context more amenable to tokenization in the next iteration. However, this approach is sub-optimal to resolving irregular tokens and we leave a systematic investigation as future work.

\paragraph{Fallback for Infilling Failures.}
Sometimes self-infilling does not yield a coherent generation, such as failing to generate a well-defined suffix or producing an infilled middle without joining the suffix appropriately. To mitigate this issue and continue the looping mechanism, we employ a simple fallback strategy to reset the context when necessary. For self-infilling, if it fails to generate a proper suffix, we retain the prefix in the context and pass it to the subsequent left-to-right decoding; alternatively, if the middle fails to integrate with the given suffix, we truncate it to the last occurrence of the suffix prompt and use the resulting segment for left-to-right generation. Similarly, if the left-to-right decoding leads to degenerate outputs or fails to produce suffix prompt tokens, we revert to the context of the previous self-infilling call. Such fallback strategies ensure that even though some updating iterations encounter issues, we can still reset the context properly and continue the looping mechanism.

\paragraph{Suffix Prompts.}
\begin{figure}[ht!]
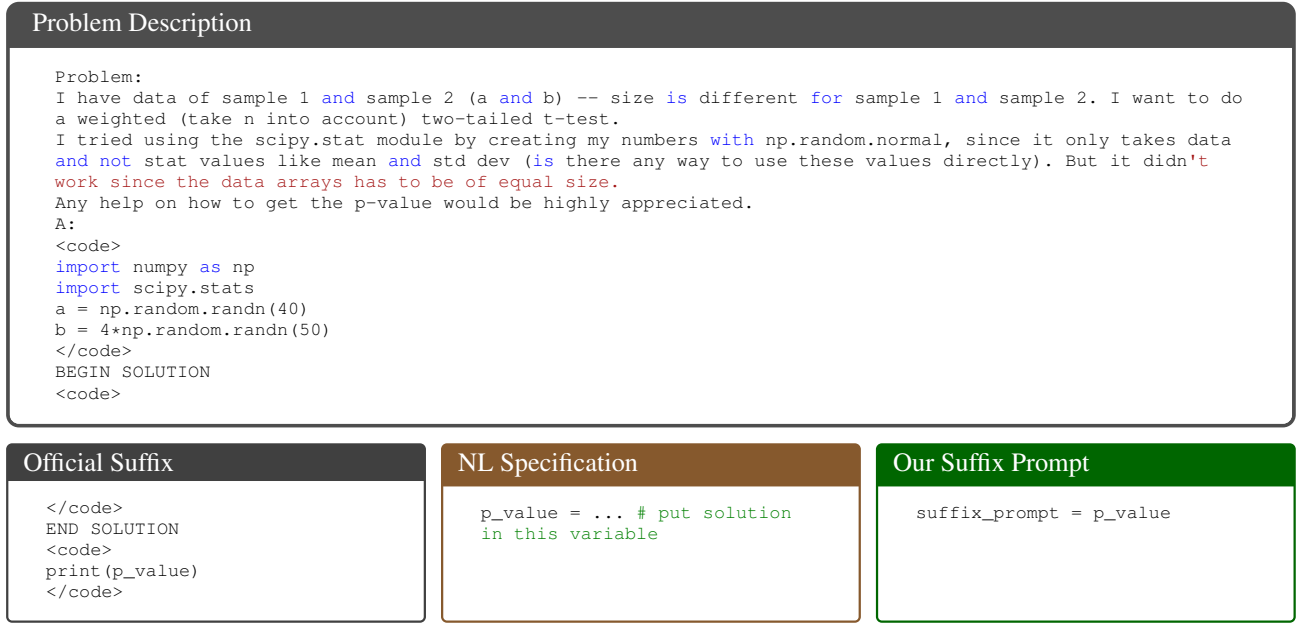

\begin{DSOuterBox}{Problem Description}
\begin{codelisting}{python}
Problem:
I have data of sample 1 and sample 2 (a and b) -- size is different for sample 1 and sample 2. I want to do a weighted (take n into account) two-tailed t-test.
I tried using the scipy.stat module by creating my numbers with np.random.normal, since it only takes data and not stat values like mean and std dev (is there any way to use these values directly). But it didn't work since the data arrays has to be of equal size.
Any help on how to get the p-value would be highly appreciated.
A:
<code>
import numpy as np
import scipy.stats
a = np.random.randn(40)
b = 4*np.random.randn(50)
</code>
BEGIN SOLUTION
<code>
\end{codelisting}
\end{DSOuterBox}
\begin{tcbitemize}[raster columns=3, raster equal height=rows, size=small]
\tcbitem[title=Official Suffix,colback=white]
\begin{codelisting}{python}
</code>
END SOLUTION
<code>
print(p_value)
</code>
\end{codelisting}
\tcbitem[title=NL Specification,colback=white,colframe=brown!70!black]
\begin{codelisting}{python}
p_value = ... # put solution in this variable
\end{codelisting}
\tcbitem[title=Our Suffix Prompt,colback=white,colframe=green!40!black]
\begin{codelisting}{python}
suffix_prompt = p_value
\end{codelisting}
\end{tcbitemize}
\caption{Illustration of suffix prompt construction for \dsonek problems \citep{lai2023ds1000}, where the official suffix for insertion mode is manually translated to a natural language (NL) specification and appended to the completion mode prompt. Self-infilling generation follows the completion format and utilizes the suffix prompt derived from the NL specification.}
\label{fig:example:sp_for_ds1000}
\end{figure}
For \humaneval, \multiple, \mbpp, and \gsm benchmarks, which focus on function-level program synthesis, we set the default suffix prompt to \mintinline{python}{return}, controlling the structure of the generation while relying on minimal prior knowledge.

The \dsonek benchmark covers a spectrum of code generation tasks. We adopt a simple heuristic to construct suffix prompts according to the provided input/output specification in a problem-wise manner. In particular, \dsonek offers two prompt formats for evaluating code language models: \textbf{1)} insertion mode, where official prefixes and suffixes are given as input for each problem, except in Matplotlib tasks where a trailing context is absent; and \textbf{2)} left-to-right mode with only prefix prompts and manually translated natural language (NL) specifications for suffixes appended to prefixes. To evaluate the ability of self-infilling, we follow the left-to-right mode and devise suffix prompts from the translated NL specification for each problem, as illustrated in \cref{fig:example:sp_for_ds1000}. In general, our strategy proceeds as follows: if the problem requires completing a Python function, we set $\psp = ``\texttt{return}"$; if the problem requires storing the result in a particular variable named $\texttt{var}$, set $\psp = ``\texttt{var}"$\footnote{If there are multiple variables of interest, we simply adopt the last variable name as the suffix prompt.}; if there is no such specification, we set $\psp = ``\texttt{</code>}"$ that signifies the end of the generation; as for Matplotlib problems, we set $\psp = ``\texttt{\# SOLUTION END}"$ that indicates the end.

\subsection{Evaluation Setup Details}
\label{app:additional_exp_details:eval}
\paragraph{Stop Criterion.}
Code language models usually require a list of stop tokens to signify the end of their generation. For \humaneval, we recruit the same set of stop tokens following prior research \citep{chen2021codex,li2023starcoder}, including ["\verb$\nclass$", "\verb$\ndef$", "\verb$\n#$", "\verb$\n@$", "\verb$\nprint$", "\verb$\nif$", "\verb$\n```$"] that indicates the start of generation beyond the current function scope. Following a similar spirit, ["\verb$\nclass$", "\verb$\nassert$", '\verb$\n"""$', "\verb$\nprint$", "\verb$\nif$", "\verb$\n<|/$", "\verb$\n```$", "\verb$[DONE]$"] is used for \mbpp. Default stop tokens are used for \multiple \citep{cassano2022multiple}. For \dsonek, the default setup in \citet{lai2023ds1000} adopts stop tokens ["\verb$</code>$", "\verb$# SOLUTION END$"]. We extend the list to include ["\verb$</code>$", "\verb$# SOLUTION END$", "\verb$\nEND SOLUTION$"], which more effectively truncates generated outputs and improves performance across both baselines and our method. For \gsm, which employs a few-shot prompting format, the stop tokens are ["\verb$\n\n\n$", "\verb$\nQ:$"].

\paragraph{On Evaluating Looping with Pass@$k$ Metrics.}
When evaluating pass@$k$ metrics for code generation, it is worth noting that the looping mechanism generates a single solution within an updated context iteratively, unlike multi-pass approaches where each pass yields an independent solution for subsequent selection and evaluation. The looping process is essentially part of a singular generative process and does not perform interim evaluations of functional correctness or involve selecting among different iterations. Consequently, our evaluation of pass@$k$ metrics still accurately reflects functional correctness without skewing or biasing the results due to the looping process. This ensures our comparison to conventional left-to-right decoding on Pass@$k$ remains valid.

\section{Additional Experimental Results}
\label{app:additional_exp_results}

\subsection{Mathematical Reasoning with \gsm}
\label{app:additional_exp_results:gsm8k}
In this section, we extend our analysis to the \gsm \citep{cobbe2021gsm} mathematical reasoning task, employing methodologies from Program-Aided Language models \citep[PAL;][]{gao2023pal} and Chain-of-Thought prompting \citep[CoT;][]{wei2022cot}. PAL solves mathematical reasoning problems by generating and executing Python programs to calculate answers, whereas CoT generates explicit intermediate steps in natural language to reach conclusions. For self-infilling in PAL, we set the default suffix prompt to \mintinline{python}{return}; while for CoT, we design the suffix prompt as follows, 
\begin{align*}
    \psp \coloneqq ``\texttt{\textbackslash nTherefore, the answer is}".
\end{align*} 
This formulation guides the suffix towards generating a conclusive response to the mathematical problem.

\gsm results are presented in \cref{tb:gsm8k}. We observe a consistent improvement in reasoning accuracy over left-to-right baselines with \codellama models under the PAL format. However, similar to the \dsonek benchmark, the benefits for \starcoder models are slight. In terms of CoT, which involves natural language instead of code, self-infilling decoding also leads to improved task accuracy in comparison to traditional left-to-right generation. These results not only highlight the versatility of self-infilling but also suggest its potential applicability in complicated reasoning tasks.

\begin{table}[t]
\caption{8-shot accuracy on the \gsm math-reasoning benchmark. Solutions are generated via chain-of-thought prompting \citep{wei2022cot,kojima2022zeroshotcot} with greedy decoding. \textsuperscript{\textdagger} Results are taken from \citet{li2023starcoder,touvron2023llama,roziere2023codellama}.}
\label{tb:gsm8k}
\vskip 0.1in
\centering
\resizebox{0.6\linewidth}{!}{
\begin{tabular}{rrrcc}
\toprule 
\textbf{Model} & \textbf{Size} & \textbf{Method} & \textbf{GSM8K PAL} & \textbf{GSM8K CoT} \\
\midrule
CodeGen-Multi\textsuperscript{\textdagger} & 16B & \multirow{3}{*}{Left-to-right} & 8.6 & 3.2 \\
CodeGen-Mono\textsuperscript{\textdagger} & 16B & & 13.1 & 2.6 \\
\starcoderbase\textsuperscript{\textdagger} & 15.5B & & 21.5 & 8.4 \\
\midrule
\midrule
\multirow{2}{*}{\starcoderbase} & \multirow{2}{*}{15.5B}
& Left-to-right & 21.5 & 6.1 \\ 
& & Self-infilling & 21.3 & 6.4 \\
\midrule
\multirow{2}{*}{\starcoder} & \multirow{2}{*}{15.5B}
& Left-to-right & 23.9 & 5.8 \\
& & Self-infilling & 24.9 & 7.0 \\ 
\midrule
\multirow{2}{*}{\codellama} & \multirow{2}{*}{7B}
& Left-to-right & 27.4 & 9.7 \\
& & Self-infilling & 29.4 & 10.2 \\ 
\midrule
\multirow{2}{*}{\codellama} & \multirow{2}{*}{13B}
& Left-to-right & 37.7 & 17.7 \\
& & Self-infilling & 39.8 & 21.6 \\ 
\bottomrule
\end{tabular}}
\end{table}

\begin{wraptable}[12]{R}{0.5\columnwidth}
\vspace{-0.2in}
\caption{Results on \humaneval and \dsonek with different suffix prompts.}
\label{app:tb:ablation:sp}
\vspace{-0.25in}
\begin{center}
\resizebox{0.5\columnwidth}{!}{ 
\begin{tabular}[t]{ll|cccc}
\toprule
\multirow{3}{*}{Dataset} & \multirow{3}{*}{Suffix Prompt} & \multicolumn{4}{c}{Model} \\
& & \multicolumn{2}{c}{\starcoder} & \multicolumn{2}{c}{\codellama} \\
& & 15B-\textsc{Base} & 15B & 7B & 13B \\
\midrule
\multirow{4}{*}{\humaneval}
& "" & 29.9 & 34.1 & 33.5 & 35.4 \\
& \texttt{"return"} (\textbf{default}) & 33.5 & 37.8 & 34.1 & 38.4 \\
& \texttt{"\$ARG\_NAME"} & 26.8 & 31.1 & 29.2 & 37.8 \\
& \texttt{"\textbackslash nreturn result\textbackslash n"} & 32.9 & 38.4 & 30.5 & 37.2 \\
\midrule
\multirow{3}{*}{\dsonek} 
& "" & 27.0 & 28.6 & 24.7 & 27.6\\
& \texttt{"\$ARG\_NAME"} (\textbf{default}) & 27.0 & 29.9 & 28.1 & 31.6 \\
& \texttt{"\textbackslash n</code>\textbackslash n"} & 26.9 & 30.0 & 23.6 & 27.1 \\
\bottomrule
\end{tabular}}  
\end{center}
\end{wraptable}
\subsection{On the Effect of Different Suffix Prompts}
\label{app:additional_exp_results:different_sp}
Suffix prompts serve as another core component in our self-infilling framework, significantly influencing the decoding process and the structure of generation. To understand their effect, we conduct an experiment with varied suffix prompt configurations, as detailed in \cref{app:tb:ablation:sp}. For these experiments, we use $N=1$ for the looped mechanism to facilitate a direct comparison among different suffix prompts. For \humaneval, we explore four suffix prompt variants, including 1) an empty string, 2) the default choice with the \mintinline{python}{return} keyword, 3) a dynamic strategy \texttt{\$ARG\_NAME} using the function's first argument name, as well as 4) a full return statement by specifying a variable \mintinline{python}{result}. We observe that the default suffix prompt choice yields the best performance across all model types, but choices 3) and 4) also demonstrate effectiveness. This highlights the potential benefits of enhancing self-infilling through refined suffix prompt design. We conduct a similar evaluation scheme for \dsonek and find \starcoder models are robust to varying suffix prompts, while \codellama models benefit from more informative suffix prompts. Interestingly, \codellama models tend to produce empty suffixes with an empty suffix prompt, likely a consequence of the pre-training phase, where \texttt{<SUF>} might be commonly followed by \texttt{<MID>} and thus empty suffixes are preferred.

\begin{table}[t]
\caption{Pass@1(\%) Results on \humaneval using different looping mechanisms. $N$ denotes the number of iterations and $t$ denotes the temperature for rejection sampling. SI denotes Self-Infilling.}
\label{app:tb:ablation:loop_wo_si}
\vspace{0.1in}
\begin{center}
\resizebox{0.45\linewidth}{!}{
\begin{tabular}{lc|cccc}
\toprule
\multirow{3}{*}{Decoding Method} & \multirow{3}{*}{Setup} & \multicolumn{4}{c}{Model} \\
& & \multicolumn{2}{c}{\starcoder} & \multicolumn{2}{c}{\codellama} \\
& & 15B-\textsc{Base} & 15B & 7B & 13B \\
\midrule
\multirow{3}{*}{Rejection Sampling}
& $t\!=\!0.3$ & 32.3 & 35.4 & 34.8 & 37.8 \\
& $t\!=\!0.5$ & 31.7 & 35.4 & 35.4 & 36.6 \\
& $t\!=\!0.8$ & 32.9 & 37.2 & 34.1 & 38.4 \\
\midrule
\multirow{3}{*}{Loop w/o SI}
& $N\!=\!0$ & 31.7 & 35.4 & 34.1 & 35.4 \\
& $N\!=\!1$ & 31.1 & 34.1 & 31.7 & 34.8 \\
& $N\!=\!2$ & 31.7 & 36.0 & 31.1 & 32.9 \\
\midrule
\multirow{3}{*}{Loop w/ SI (ours)}
& $N\!=\!0$ & 27.4 & 29.2 & 29.9 & 32.3 \\
& $N\!=\!1$ & 33.5 & 37.8 & 34.1 & 38.4 \\
& $N\!=\!2$ & 36.0 & 38.4 & 39.0 & 40.8 \\
\midrule
Left-to-right & - & 31.7 & 35.4 & 34.1 & 35.4 \\
\bottomrule
\end{tabular}}  
\end{center}
\end{table}

\subsection{A Looping Mechanism without Self-infilling}
\label{app:additional_exp_results:loop_wo_si}
In this section, we investigate the impact of removing the self-infilling component from the looping mechanism. In particular, we design a looping mechanism without self-infilling by initiating with left-to-right decoding, manually extracting the latter part of the completion as \psuffix, and then using the standard infilling operator to regenerate \pmiddle based on the extracted \psuffix. This roughly corresponds to a right-shifted version of looping in \cref{model:self_infilling:looped} (instead of starting with self-infilling and then performing left-to-right decoding, this variant starts from left-to-right decoding to self-infilling). As shown in \cref{app:tb:ablation:loop_wo_si}, our findings indicate that looping under these conditions typically does not enhance coding performance, and extending the looping time does not yield improvements. Furthermore, this looping mechanism even leads to lower pass rates sometimes. This may be attributed to the specificity of \psuffix generated through left-to-right decoding, which potentially restricts the possible outcomes of \pmiddle. In addition, left-to-right decoding does not regularize the structure of generated output, which might be degenerate and thus lead to degenerate \psuffix as well. 
Self-infilling, on the other hand, effectively scaffolds the overall generation and regenerates \psuffix based on a succinct \psp, encouraging the model to explore a larger decoding space. This ablation study underscores the value of the self-infilling component in the looping mechanism.

We also compare our approach to another iterative variant with rejection sampling. This method initially uses greedy decoding and falls back to stochastic sampling if the initial generation degenerates (that is, lacking a return keyword or becoming empty). We limit the rejection process to a maximum of 10 trials to prevent infinite loops. The temperature settings were varied, with top\_p set at 0.95. The results indicate that while rejection sampling reduces degenerate behaviors (typically requiring 2-3 sampling attempts for acceptance), it does not significantly improve pass rates compared to self-infilling. Additionally, we did not find a significant correlation between pass rates, sampling times, and temperature settings. This evidence suggests that the advantages of self-infilling extend beyond simply addressing degeneration biases.

\subsection{Comparison to Sample-and-Rank Baselines}
\label{app:additional_exp_results:comp_to_multi_sample}

The introduced looping mechanism (\cref{model:self_infilling:looped}) in place updates pieces of the generation multiple times, which produces multiple possible generations along the iterative process. An alternative approach to accomplishing this is performing beam search, or simply sampling multiple generations in the conventional left-to-right way, followed by selection based on the highest probability or average token log-likelihood \citep{chen2021codex,zhang2023coder-ranker}. Our comparative analysis with these multi-sample baselines, as presented in \cref{tb:ablation:multi_sample} on \humaneval, reveals that our looping mechanism, despite lacking an explicit selection or aggregation step, performs competitively with these baselines. Furthermore, when complemented by a ranking component --- first generating multiple samples via self-infilling and then selecting based on the highest average likelihood --- our method demonstrates superior performance over existing baselines. These results highlight the effectiveness of the looping mechanism, which encourages code language models to explore a broader decoding space.

Our looping mechanism is also relevant to multi-pass code generation approaches in the literature \citep{chen2023selfdebug,jiang2023selfevolve,pan2023selfcorrection_survey}. Unlike these approaches \citet{chen2023selfdebug}, we demonstrate that base code models, even without instruction-following abilities, can exhibit improved generation quality via looping. Our approach alternates between left-to-right decoding and self-infilling, relying solely on the pre-trained code model without additional training. Nevertheless, our work can be further augmented with other signals like syntactic representation \citep{zheng2023chaincoder} and execution results \citep{ni2023lever} to generate better samples.

\subsection{Additional Ablation Studies}
\label{app:additional_exp_results:ablation_tau_n}
\begin{figure}[t]
\centering
\includegraphics[width=0.75\textwidth]{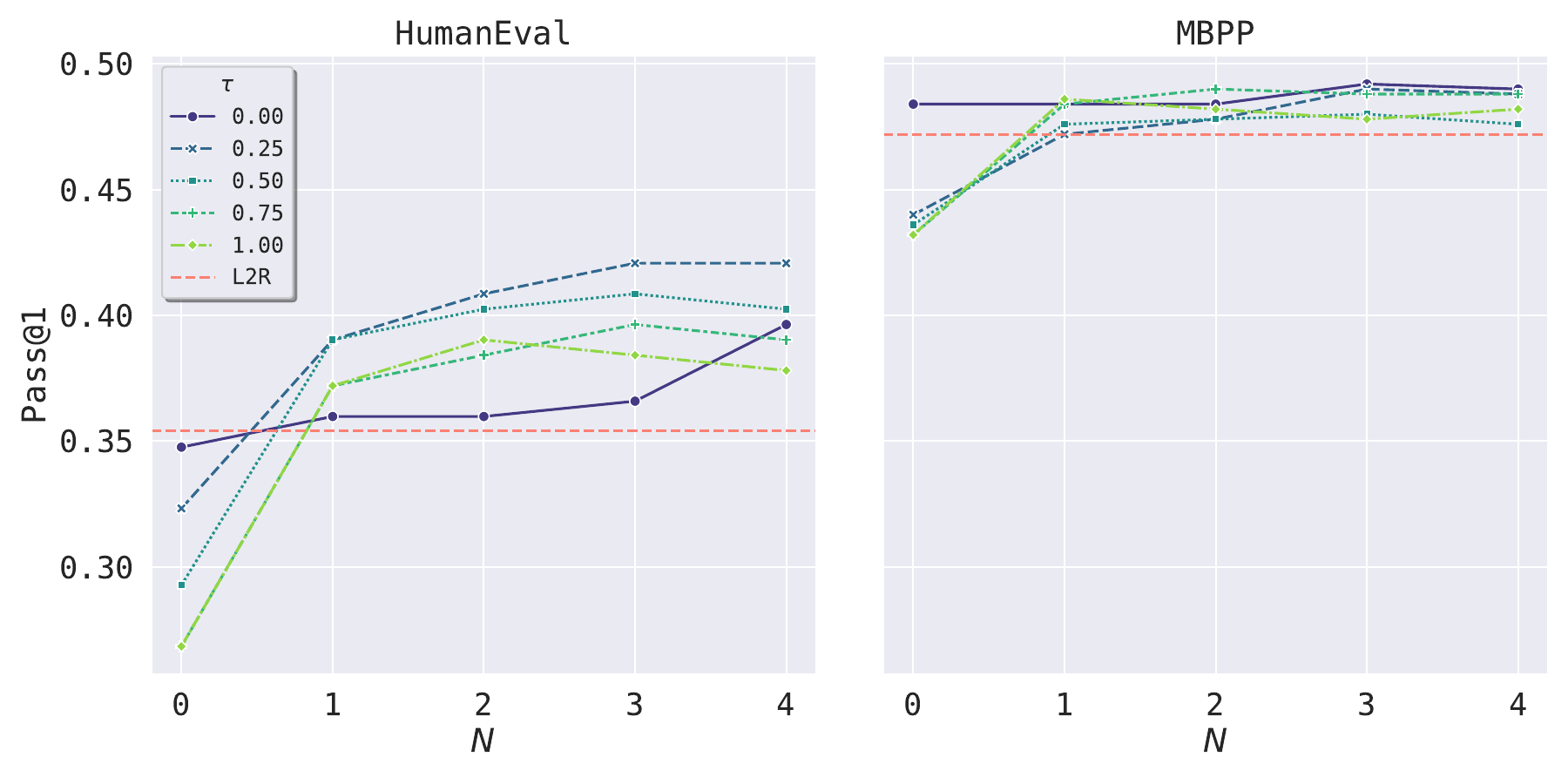}
\caption{Results on \humaneval and \mbpp with different values of probability threshold $\tau$ and looping times $N$ on \codellama~13B. $N=0$ indicates the looping mechanism is disabled, and the horizontal dashed line represents the performance of the vanilla left-to-right baseline (L2R).}
\label{fig:ablation:tau_n_codellama13b}
\end{figure}

\begin{figure}[t]
\centering
\includegraphics[width=0.75\textwidth]{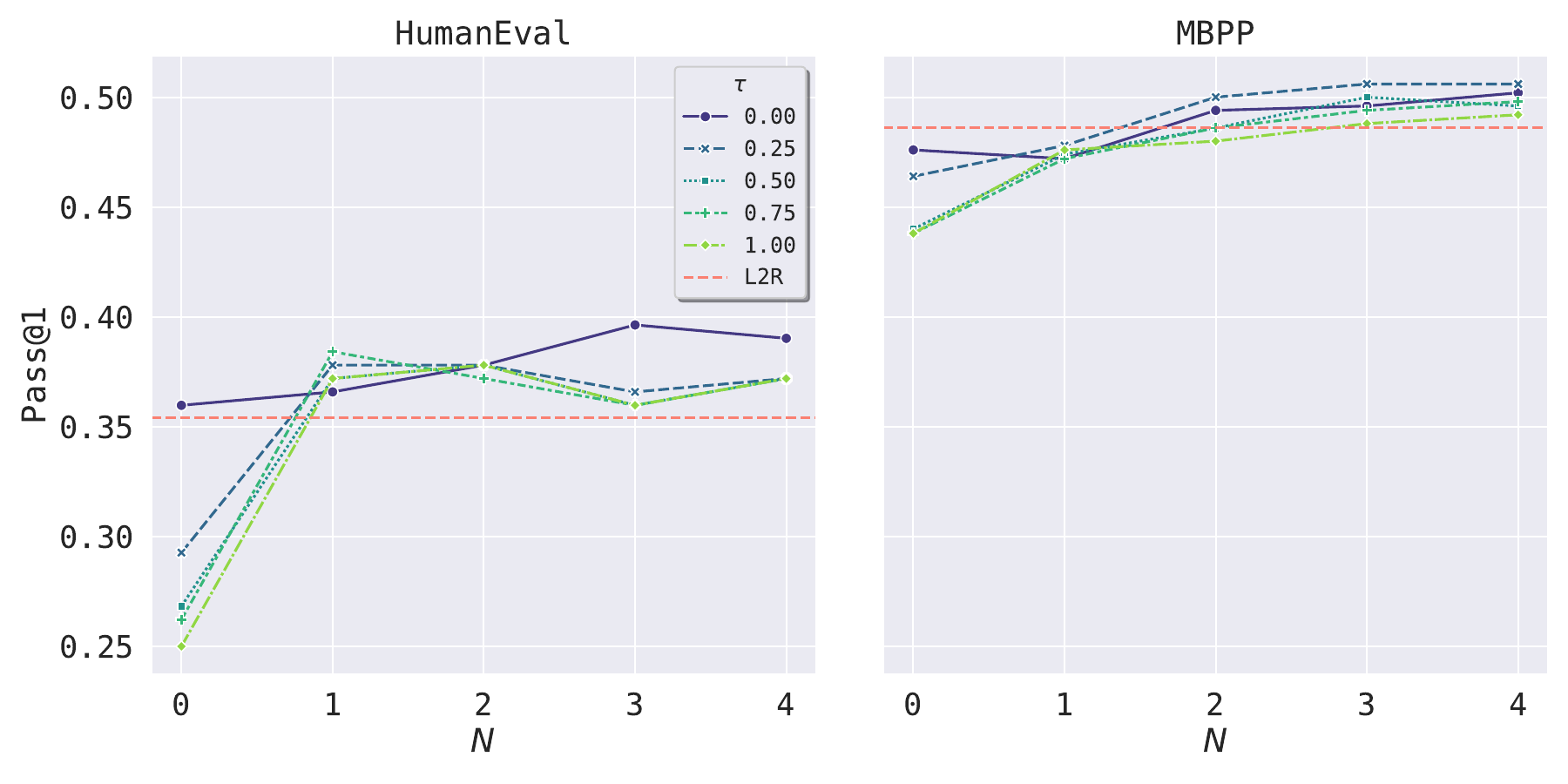}
\caption{Results on \humaneval and \mbpp with different values of probability threshold $\tau$ and looping times $N$ on \starcoder. $N=0$ indicates that the looping mechanism is disabled, and the horizontal dashed line represents the performance of the vanilla left-to-right baseline (L2R).}
\label{fig:ablation:tau_n_starcoder}
\end{figure}

\begin{figure}[t]
\centering
\includegraphics[width=0.75\textwidth]{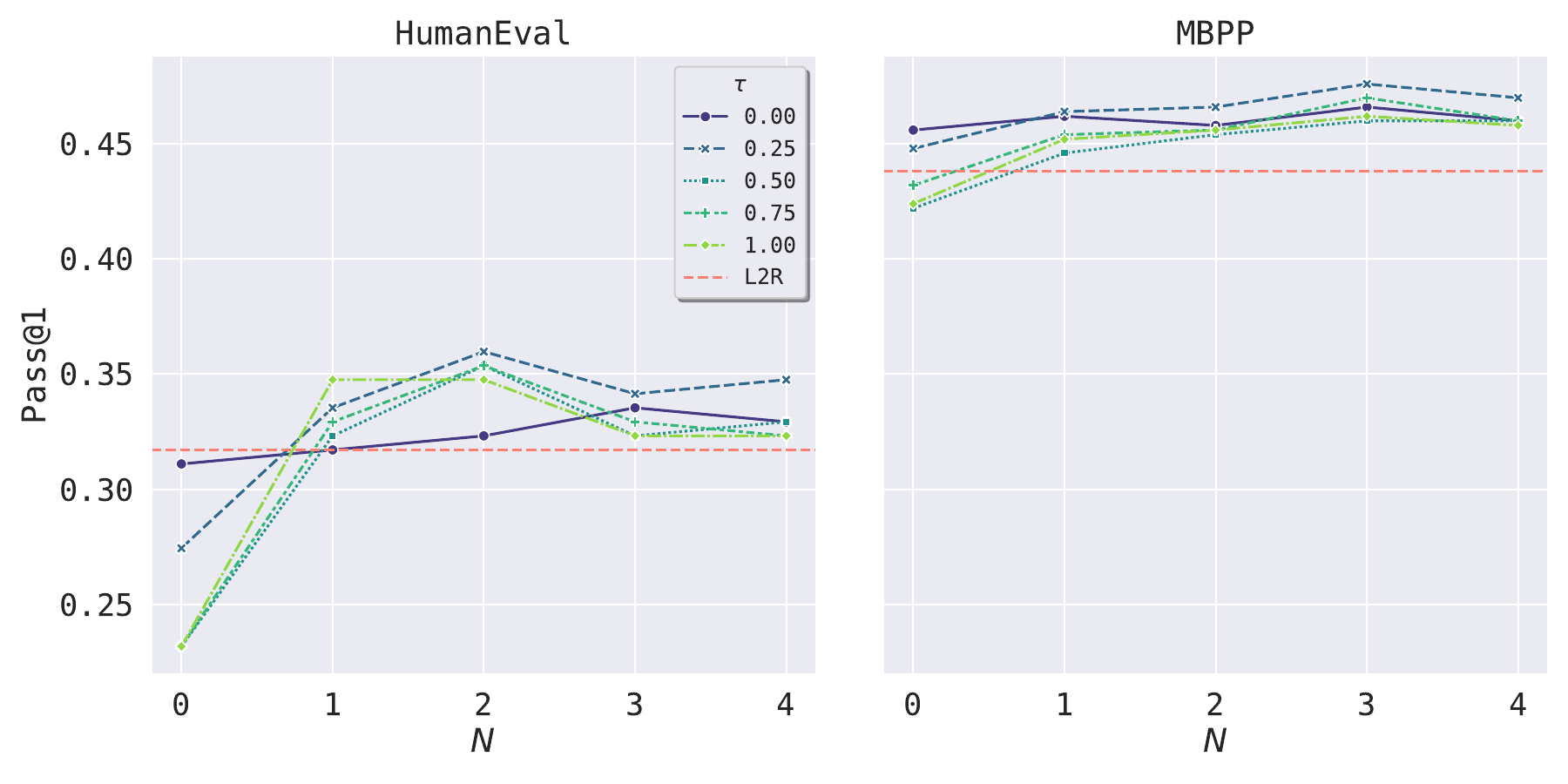}
\caption{Results on \humaneval and \mbpp with different values of probability threshold $\tau$ and looping times $N$ on \starcoderbase. $N=0$ indicates that the looping mechanism is disabled, and the horizontal dashed line represents the performance of the vanilla left-to-right baseline (L2R).}
\label{fig:ablation:tau_n_starcoderbase}
\end{figure}

This section provides additional ablation studies on the effect of hyper-parameters $\tau$ and $N$ under various code language models, including \codellama~13B (\cref{fig:ablation:tau_n_codellama13b}), \starcoder (\cref{fig:ablation:tau_n_starcoder}), and \starcoderbase (\cref{fig:ablation:tau_n_starcoderbase}). Note that setting $\tau = 0$ and $N=0$ reduces the looping mechanism to conventional left-to-right decoding. However, this configuration still adds a sentinel \texttt{<PRE>} to the beginning of the input prompt, potentially leading to a subtle performance difference.

\subsection{Examples of Self-infilling}
\label{app:additional_exp_results:examples}
\begin{figure}[h!]
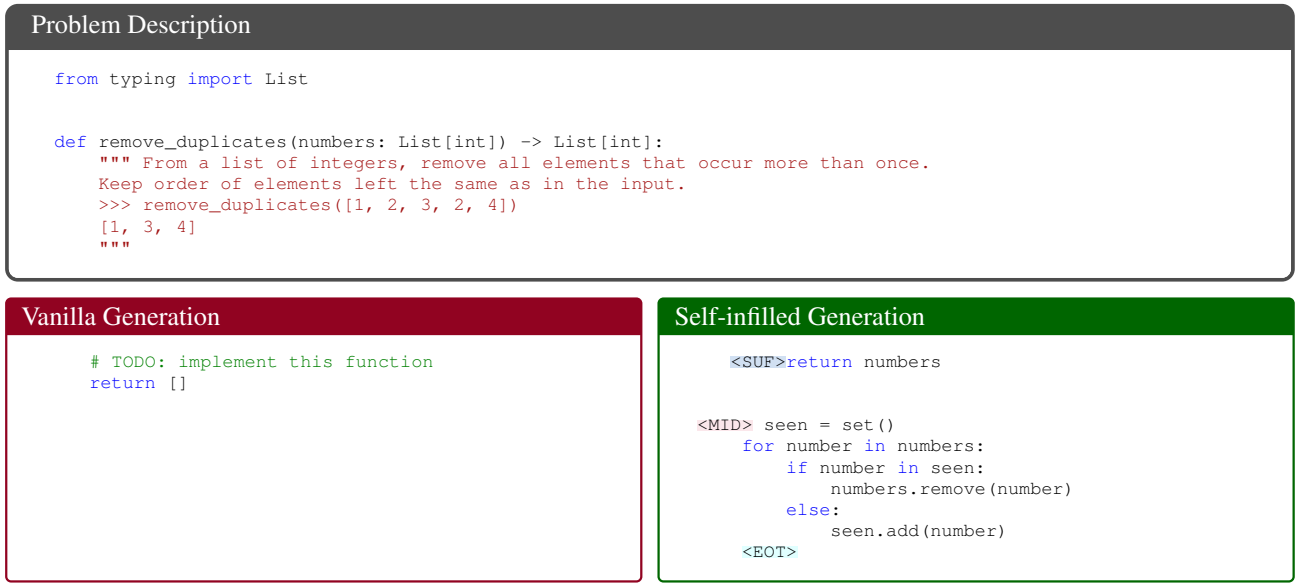

\begin{DSOuterBox}{Problem Description}
\begin{codelisting}{python}
from typing import List


def remove_duplicates(numbers: List[int]) -> List[int]:
    """ From a list of integers, remove all elements that occur more than once.
    Keep order of elements left the same as in the input.
    >>> remove_duplicates([1, 2, 3, 2, 4])
    [1, 3, 4]
    """
\end{codelisting}
\end{DSOuterBox}
\begin{tcbitemize}[raster columns=2, raster equal height=rows, size=small]
\tcbitem[title=Vanilla Generation,colback=white,colframe=lava]
\begin{codelisting}{python}
    # TODO: implement this function
    return []
\end{codelisting}
\tcbitem[title=Self-infilled Generation,colback=white,colframe=green!40!black]
\begin{codelisting}{python}
   |\codecolorbox{suf}{<SUF>}|return numbers


|\codecolorbox{mid}{<MID>}| seen = set()
    for number in numbers:
        if number in seen:
            numbers.remove(number)
        else:
            seen.add(number)
    |\codecolorbox{eot}{<EOT>}|
\end{codelisting}
\end{tcbitemize}
\caption{An illustration of self-infilling decoding with \textbf{interruption} on \humaneval: vanilla left-to-right generation exhibits degenerate behaviors that output an empty program due to the initially generated comment; however, self-infilling early interrupts the decoding flow and drafts a plausible suffix first, which then drives subsequent decoding towards joining that suffix.}
\label{fig:example:l2r_vs_si:1}
\end{figure}

\begin{figure}[h!]
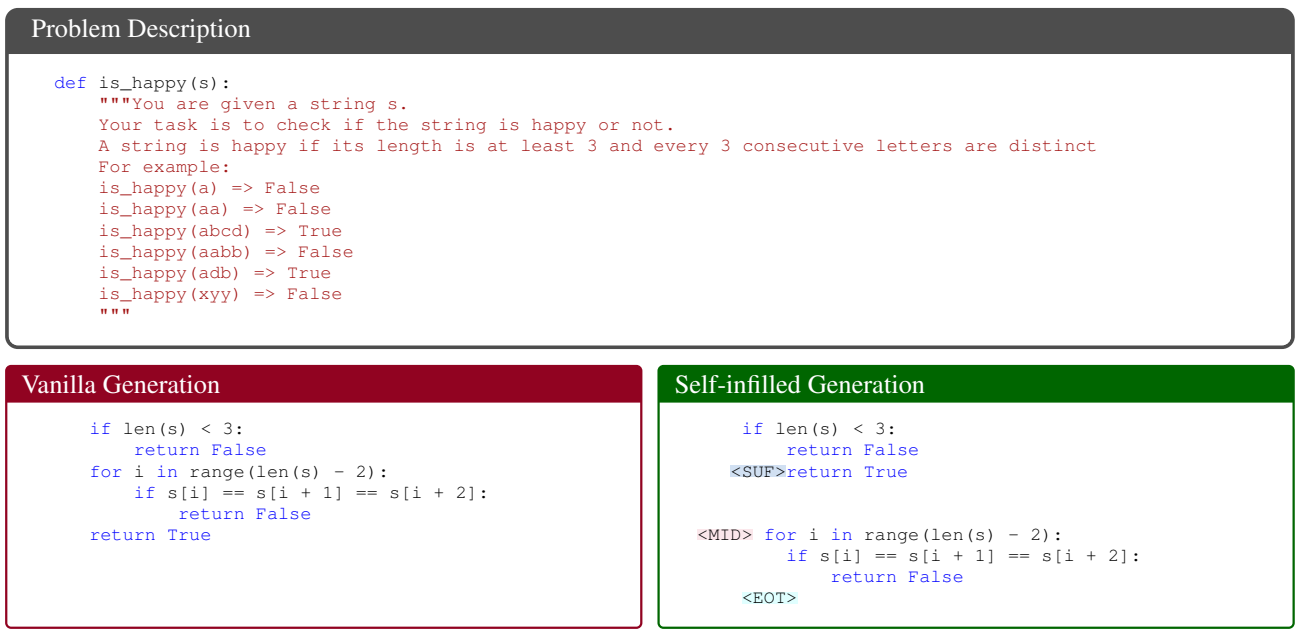

\begin{DSOuterBox}{Problem Description}
\begin{codelisting}{python}
def is_happy(s):
    """You are given a string s.
    Your task is to check if the string is happy or not.
    A string is happy if its length is at least 3 and every 3 consecutive letters are distinct
    For example:
    is_happy(a) => False
    is_happy(aa) => False
    is_happy(abcd) => True
    is_happy(aabb) => False
    is_happy(adb) => True
    is_happy(xyy) => False
    """
\end{codelisting}
\end{DSOuterBox}
\begin{tcbitemize}[raster columns=2, raster equal height=rows, size=small]
\tcbitem[title=Vanilla Generation,colback=white,colframe=lava]
\begin{codelisting}{python}
    if len(s) < 3:
        return False
    for i in range(len(s) - 2):
        if s[i] == s[i + 1] == s[i + 2]:
            return False
    return True
\end{codelisting}
\tcbitem[title=Self-infilled Generation,colback=white,colframe=green!40!black]
\begin{codelisting}{python}
    if len(s) < 3:
        return False
   |\codecolorbox{suf}{<SUF>}|return True


|\codecolorbox{mid}{<MID>}| for i in range(len(s) - 2):
        if s[i] == s[i + 1] == s[i + 2]:
            return False
    |\codecolorbox{eot}{<EOT>}|
\end{codelisting}
\end{tcbitemize}
\caption{An illustration of self-infilling decoding with \textbf{interruption} on \humaneval: both vanilla left-to-right and self-infilling generation produce the same code snippet, despite through different decoding trajectories.}
\label{fig:example:l2r_vs_si:2}
\end{figure}

In this section, we present additional examples (\Cref{fig:example:l2r_vs_si:1,,fig:example:l2r_vs_si:2,,fig:example:l2r_vs_si:3,,fig:example:l2r_vs_si:4} for \humaneval and \Cref{fig:example:ds1000:1,,fig:example:ds1000:2,,fig:example:ds1000:3} for \dsonek problems) to illustrate the distinct decoding behaviors of the self-infilling approach compared to traditional left-to-right decoding. Besides, we also provide demonstrations of self-infilling generation with the interruption and looping mechanism, as shown in \Cref{fig:example:loop:correct_to_wrong,,fig:example:loop:remain_incorrect,,fig:example:loop:remain_correct,,fig:example:loop:unchanged,,fig:example:loop:another_wrong_to_correct,,fig:example:loop:wrong_to_correct}.

These examples are generated using \codellama~13B with greedy decoding. For the sake of brevity and clarity, the occurrence of stop tokens during generation is omitted. Also note that the sentinel token \texttt{<PRE>} used in self-infilling, which is usually positioned at the beginning of the problem description, is not displayed in these illustrations for readability.

\begin{table}[t]
\caption{Comparison results of self-infilling against various multi-sample baselines on \humaneval. $N$ denotes the number of iterations in our looping mechanism, while $S$ indicates the beam size for Beamsearch and the number of samples for other approaches. Samples are generated with nucleus sampling at temperature 0.3 and top-$p$ 0.95. The \emph{Rank} strategy selects the generation with the highest mean token log probabilities \citep{chen2021codex}.}
\label{tb:ablation:multi_sample}
\vskip 0.1in
\begin{center}
\resizebox{0.5\linewidth}{!}{ 
\begin{tabular}{lc|cccc}
\toprule
\multirow{3}{*}{Method} & \multirow{3}{*}{Setup} & \multicolumn{4}{c}{Model} \\
& & \multicolumn{2}{c}{\starcoder} & \multicolumn{2}{c}{\codellama} \\
& & 15B-\textsc{Base} & 15B & 7B & 13B \\
\midrule
Greedy & $S=1$ &  31.7  & 35.4   & 34.1   &  35.4  \\
\midrule
\multirow{2}{*}{Beamsearch} & $S=2$ &  36.0  & 37.2   & 34.1   &  40.8  \\
& $S=4$ &  34.8  & \textbf{40.2}  &  16.5  &   19.5 \\
\midrule
Sample & $S=1$ &  29.3  & 30.5   & 29.3   &  34.8  \\
\midrule
\multirow{2}{*}{Rank}  & $S=2$ &  36.0  & 37.2   & 37.2   &  37.2  \\
& $S=4$ &  32.9  & 36.6   & 32.3   &  34.1  \\
\midrule
\midrule
\multirow{2}{*}{Ours} & $S=1,N=1$ &  33.5  & 37.8   & 34.1   &  38.4  \\
& $S=1,N=2$ &  36.0  & 38.4   & \textbf{39.0}  &   40.8 \\
\midrule
\multirow{2}{*}{Ours + Rank} & $S=2,N=1$ &  34.8  & \textbf{40.2}  &  33.5  & \textbf{42.1} \\
& $S=2,N=2$ &  \textbf{37.8} &  37.8  &  36.5  &   40.2 \\
\bottomrule
\end{tabular}}  
\end{center}
\end{table}

\newpage

\begin{figure}[h!]
\begin{DSOuterBox}{Problem Description}
\begin{codelisting}{python}
def total_match(lst1, lst2):
    '''
    Write a function that accepts two lists of strings and returns the list that has 
    total number of chars in the all strings of the list less than the other list.

    if the two lists have the same number of chars, return the first list.

    Examples
    total_match([], []) -> []
    total_match(['hi', 'admin'], ['hI', 'Hi']) -> ['hI', 'Hi']
    total_match(['hi', 'admin'], ['hi', 'hi', 'admin', 'project']) -> ['hi', 'admin']
    total_match(['hi', 'admin'], ['hI', 'hi', 'hi']) -> ['hI', 'hi', 'hi']
    total_match(['4'], ['1', '2', '3', '4', '5']) -> ['4']
    '''
\end{codelisting}
\end{DSOuterBox}
\begin{tcbitemize}[raster columns=2, raster equal height=rows, size=small]
\tcbitem[title=Vanilla Generation,colback=white,colframe=lava]
\begin{codelisting}{python}
    # lst1 = ['hi', 'admin']
    # lst2 = ['hI', 'Hi']
    # lst1 = ['hi', 'admin']
    # lst2 = ['hi', 'hi', 'admin', 'project']
    # lst1 = ['hi', 'admin']
    # lst2 = ['hI', 'hi', 'hi']
    # lst1 = ['4']
    # lst2 = ['1', '2', '3', '4', '5']
    # lst1 = ['hi', 'admin']
    # lst2 = ['hI', 'hi', 'hi']
    # lst1 = ['hi', 'admin']
    # lst2 = ['hI', 'hi', 'hi']
    # lst1 = ['hi', 'admin']
    # lst2 = ['hI', 'hi', 'hi']
    # lst1 = ['hi', 'admin']
    ...
\end{codelisting}
\tcbitem[title=Self-infilled Generation,colback=white,colframe=green!40!black]
\begin{codelisting}{python}
    #|\codecolorbox{suf}{<SUF>}|return lst1
    else:
        return lst2


|\codecolorbox{mid}{<MID>}| Write your code here.
    total_chars_lst1 = 0
    total_chars_lst2 = 0
    for i in lst1:
        total_chars_lst1 += len(i)
    for i in lst2:
        total_chars_lst2 += len(i)
    if total_chars_lst1 < total_chars_lst2:
        |\codecolorbox{eot}{<EOT>}|
\end{codelisting}
\end{tcbitemize}
\caption{An illustration of self-infilling decoding with \textbf{interruption} on \humaneval: vanilla left-to-right generation tends to repeat the same snippet; self-infilling first starts the program with a symbol \texttt{\#} that signifies a comment; however, the specific content within the comment is uncertain to the model, which thus issues an interruption of current decoding and proceeds to the suffix, circumventing a context prone of degeneracy.}
\label{fig:example:l2r_vs_si:3}
\end{figure}

\begin{figure}[h!]
\begin{DSOuterBox}{Problem Description}
\begin{codelisting}{python}
def get_closest_vowel(word):
    """You are given a word. Your task is to find the closest vowel that stands between two consonants from the right side of the word (case sensitive).
    
    Vowels in the beginning and ending doesn't count. Return empty string if you didn't find any vowel met the above condition. 

    You may assume that the given string contains English letter only.

    Example:
    get_closest_vowel("yogurt") ==> "u"
    get_closest_vowel("FULL") ==> "U"
    get_closest_vowel("quick") ==> ""
    get_closest_vowel("ab") ==> ""
    """
\end{codelisting}
\end{DSOuterBox}
\begin{tcbitemize}[raster columns=2, raster equal height=rows, size=small]
\tcbitem[title=Vanilla Generation,colback=white,colframe=lava]
\begin{codelisting}{python}
    vowels = "aeiouAEIOU"
    word = word[::-1]
    for i in range(len(word)):
        if word[i] in vowels:
            if i == 0:
                return ""
            else:
                return word[i]
    return ""
\end{codelisting}
\tcbitem[title=Self-infilled Generation,colback=white,colframe=green!40!black]
\begin{codelisting}{python}
    vowels =|\codecolorbox{suf}{<SUF>}|return vowels[0]
    else:
        return vowels[-1]


|\codecolorbox{mid}{<MID>}| "aeiouAEIOU"
    word = word[::-1]
    for i in range(len(word)):
        if word[i] in vowels:
            if i != 0:
                return word[i]
            else:
                return ""
    return ""
\end{codelisting}
\end{tcbitemize}
\caption{An illustration of self-infilling decoding with \textbf{interruption} on \humaneval: both vanilla left-to-right and self-infilling generation produce an incorrect code snippet. While self-infilling is invoked to yield a suffix first, it fails to complete the infilling operation and keeps generating the code \emph{without} respecting the previously produced suffix.}
\label{fig:example:l2r_vs_si:4}
\end{figure}

\begin{figure}[ht]
\centering
\begin{DSOuterBox}{Problem Description}
\begin{codelisting}{python}
Problem:

I have used the

sklearn.preprocessing.OneHotEncoder
to transform some data the output is scipy.sparse.csr.csr_matrix how can I merge it back into my original dataframe along with the other columns?

I tried to use pd.concat but I get

TypeError: cannot concatenate a non-NDFrame object
Thanks

A:

<code>
import pandas as pd
import numpy as np
from scipy.sparse import csr_matrix
df_origin, transform_output = load_data()
def solve(df, transform_output):
    # return the solution in this function
    # result = solve(df, transform_output)
    ### BEGIN SOLUTION
\end{codelisting}
\end{DSOuterBox}
\begin{tcbitemize}[raster columns=2, raster equal height=rows, size=small]
\tcbitem[title=Vanilla Generation,colback=white,colframe=lava]
\begin{codelisting}{python}
    # TODO: Implement your solution here
    #
    #
    #
    #
    #
    # ...
\end{codelisting}
\tcbitem[title=Self-infilled Generation,colback=white,colframe=green!40!black]
\begin{codelisting}{python}
   |\codecolorbox{suf}{<SUF>}|return result

result = solve(df_origin, transform_output)
|\codecolorbox{mid}{<MID>}| result = pd.concat([df, transform_output], axis=1)
    ### END SOLUTION
    |\codecolorbox{eot}{<EOT>}|
\end{codelisting}
\end{tcbitemize}
\caption{An illustration of self-infilling decoding with \textbf{interruption} on \dsonek: vanilla left-to-right generation exhibits degenerate behaviors that output an empty program since it produces a misleading comment at the very beginning of the solution; however, self-infilling early drafts a plausible suffix first, which in turn drives the overall generation process to yield a plausible solution.}
\label{fig:example:ds1000:1}
\end{figure}

\begin{figure}[ht]
\centering
\begin{DSOuterBox}{Problem Description}
\begin{codelisting}{python}
Problem:

I may be missing something obvious, but I can't find a way to compute this.

Given two tensors, I want to keep elements with the minimum absolute values, in each one of them as well as the sign.

I thought about

sign_x = torch.sign(x)
sign_y = torch.sign(y)
min = torch.min(torch.abs(x), torch.abs(y))
in order to eventually multiply the signs with the obtained minimums, but then I have no method to multiply the correct sign to each element that was kept and must choose one of the two tensors.


A:

<code>
import numpy as np
import pandas as pd
import torch
x, y = load_data()
</code>
signed_min = ... # put solution in this variable
BEGIN SOLUTION
<code>
\end{codelisting}
\end{DSOuterBox}
\begin{tcbitemize}[raster columns=2, raster equal height=rows, size=small]
\tcbitem[title=Vanilla Generation,colback=white,colframe=lava]
\begin{codelisting}{python}
sign_x = torch.sign(x)
sign_y = torch.sign(y)
min = torch.min(torch.abs(x), torch.abs(y))
</code>
\end{codelisting}
\tcbitem[title=Self-infilled Generation,colback=white,colframe=green!40!black]
\begin{codelisting}{python}
|\codecolorbox{suf}{<SUF>}|signed_min = torch.sign(x) * torch.min(torch.abs(x), torch.abs(y))
</code>|\codecolorbox{mid}{<MID>}||\codecolorbox{eot}{<EOT>}|
\end{codelisting}
\end{tcbitemize}
\caption{An illustration of self-infilling decoding with \textbf{interruption} on \dsonek: vanilla left-to-right generation follows the description well but does not store the result to the specified variable; instead, self-infilling ensures the overall generation must include the expected variable through suffix prompting.}
\label{fig:example:ds1000:2}
\end{figure}

\begin{figure}[ht]
\centering
\begin{DSOuterBox}{Problem Description}
\begin{codelisting}{python}
Problem:
I am trying to find col duplicates rows in a pandas dataframe.
df=pd.DataFrame(data=[[1,1,2,5],[1,3,4,1],[4,1,2,5],[5,1,4,9],[1,1,2,5]],columns=['val', 'col1','col2','3col'])
df
Out[15]: 
   val  col1  col2  3col
0    1     1     2     5
1    1     3     4     1
2    4     1     2     5
3    5     1     4     9
4    1     1     2     5
duplicate_bool = df.duplicated(subset=['col1','col2', '3col'], keep='first')
duplicate = df.loc[duplicate_bool == True]
duplicate
Out[16]: 
   val  col1  col2  3col
2    1     1     2      5
4    1     1     2      5


Is there a way to add a column referring to the index of the first duplicate (the one kept)
duplicate
Out[16]: 
   val  col1  col2 3col   index_original
2     4    1     2      5         0
4     1    1     2      5         0


Note: df could be very very big in my case....


A:
<code>
import pandas as pd


df=pd.DataFrame(data=[[1,1,2,5],[1,3,4,1],[4,1,2,5],[5,1,4,9],[1,1,2,5]],columns=['val', 'col1','col2','3col'])
</code>
result = ... # put solution in this variable
BEGIN SOLUTION
<code>
\end{codelisting}
\end{DSOuterBox}
\begin{tcbitemize}[raster columns=2, raster equal height=rows, size=small]
\tcbitem[title=Vanilla Generation,colback=white,colframe=lava]
\begin{codelisting}{python}
df['index_original'] = df.groupby(['col1','col2', '3col']).cumcount()
</code>
\end{codelisting}
\tcbitem[title=Self-infilled Generation,colback=white,colframe=green!40!black]
\begin{codelisting}{python}
|\codecolorbox{suf}{<SUF>}|result = df.loc[duplicate_bool == True]
result['index_original'] = df.loc[duplicate_bool == True].index.values
result = result.reset_index(drop=True)
</code>|\codecolorbox{mid}{<MID>}|duplicate_bool = df.duplicated(subset=['col1','col2', '3col'], keep='first')
|\codecolorbox{eot}{<EOT>}|
\end{codelisting}
\end{tcbitemize}
\caption{An illustration of self-infilling decoding with \textbf{interruption} on \dsonek: vanilla left-to-right generation adheres to the description but fails to allocate the result to the designated variable. In contrast, self-infilling guarantees that the final output is correctly assigned to the expected variable via suffix prompting.}
\label{fig:example:ds1000:3}
\end{figure}

\begin{figure}[h!]
\begin{DSOuterBox}{Problem Description}
\begin{codelisting}{python}
def get_row(lst, x):
    """
    You are given a 2 dimensional data, as a nested lists,
    which is similar to matrix, however, unlike matrices,
    each row may contain a different number of columns.
    Given lst, and integer x, find integers x in the list,
    and return list of tuples, [(x1, y1), (x2, y2) ...] such that
    each tuple is a coordinate - (row, columns), starting with 0.
    Sort coordinates initially by rows in ascending order.
    Also, sort coordinates of the row by columns in descending order.
    
    Examples:
    get_row([
      [1,2,3,4,5,6],
      [1,2,3,4,1,6],
      [1,2,3,4,5,1]
    ], 1) == [(0, 0), (1, 4), (1, 0), (2, 5), (2, 0)]
    get_row([], 1) == []
    get_row([[], [1], [1, 2, 3]], 3) == [(2, 2)]
    """
\end{codelisting}
\end{DSOuterBox}
\begin{tcbitemize}[raster columns=2, raster equal height=rows, size=small]
\tcbitem[title={$N=1:$ Self-infilling},colback=white,colframe=brown]
\tcblower
\begin{codelisting}{python}
   |\codecolorbox{suf}{<SUF>}|return []
    result = []
    for i in range(len(lst)):
        for j in range(len(lst[i])):
            if lst[i][j] == x:
                result.append((i, j))
    result.sort(key=lambda x: (x[0], -x[1]))
    return result


|\codecolorbox{mid}{<MID>}|    if not lst:
        |\codecolorbox{eot}{<EOT>}|
\end{codelisting}
\tcbitem[title={$N=1:$ Left-to-right},colback=white,colframe=bluegray]
\begin{codelisting}{python}
    if not lst:
\end{codelisting}
\tcblower
\begin{codelisting}{python}
        return []
    result = []
    for i, row in enumerate(lst):
        if x in row:
            result.append((i, row.index(x)))
    return sorted(result, key=lambda x: (x[0], -x[1]))
\end{codelisting}
\end{tcbitemize}
\begin{tcbitemize}[raster columns=2, raster equal height=rows, size=small]
\tcbitem[title={$N=2:$ Self-infilling},colback=white,colframe=brown]
\begin{codelisting}{python}
|\codecolorbox{suf}{<SUF>}|    for i, row in enumerate(lst):
        if x in row:
            result.append((i, row.index(x)))
    return
\end{codelisting}
\tcblower
\begin{codelisting}{python}
            sorted(result, key=lambda x: (x[0], -x[1]))

|\codecolorbox{mid}{<MID>}|    result = []
|\codecolorbox{eot}{<EOT>}|
\end{codelisting}
\tcbitem[title={$N=2:$ Left-to-right},colback=white,colframe=bluegray]
\begin{codelisting}{python}
    result = []
\end{codelisting}
\tcblower
\begin{codelisting}{python}
    for i, row in enumerate(lst):
        for j, col in enumerate(row):
            if col == x:
                result.append((i, j))
    return sorted(result, key=lambda x: (x[0], -x[1]))
\end{codelisting}
\end{tcbitemize}
\caption{An illustration of self-infilling decoding with \textbf{interruption} and \textbf{looping} on \humaneval, which successfully corrects the initially wrong solution at $N=1$ (classified as the `Incorrect $\rightarrow$ Correct' category in \cref{fig:loop_inspect}). The area above the dashed line indicates the current context, while the content below represents the corresponding completion.}
\label{fig:example:loop:another_wrong_to_correct}
\end{figure}

\begin{figure}[h!]
\begin{DSOuterBox}{Problem Description}
\begin{codelisting}{python}
def remove_vowels(text):
    """
    remove_vowels is a function that takes string and returns string without vowels.
    >>> remove_vowels('')
    ''
    >>> remove_vowels("abcdef\nghijklm")
    'bcdf\nghjklm'
    >>> remove_vowels('abcdef')
    'bcdf'
    >>> remove_vowels('aaaaa')
    ''
    >>> remove_vowels('aaBAA')
    'B'
    >>> remove_vowels('zbcd')
    'zbcd'
    """
\end{codelisting}
\end{DSOuterBox}
\begin{tcbitemize}[raster columns=2, raster equal height=rows, size=small]
\tcbitem[title={$N=1:$ Self-infilling},colback=white,colframe=brown]
\tcblower
\begin{codelisting}{python}
    vowels = 'aeiou'
   |\codecolorbox{suf}{<SUF>}|return text

|\codecolorbox{mid}{<MID>}| for vowel in vowels:
        text = text.replace(vowel, '')
    |\codecolorbox{eot}{<EOT>}|
\end{codelisting}
\tcbitem[title={$N=1:$ Left-to-right},colback=white,colframe=bluegray]
\begin{codelisting}{python}
    vowels = 'aeiou'
    for vowel in vowels:
        text = text.replace(vowel, '')
\end{codelisting}
\tcblower
\begin{codelisting}{python}
    return text
\end{codelisting}
\end{tcbitemize}
\begin{tcbitemize}[raster columns=2, raster equal height=rows, size=small]
\tcbitem[title={$N=2:$ Self-infilling},colback=white,colframe=brown]
\begin{codelisting}{python}
|\codecolorbox{suf}{<SUF>}|        text = text.replace(vowel, '')
    return
\end{codelisting}
\tcblower
\begin{codelisting}{python}
           text
|\codecolorbox{mid}{<MID>}|    vowels = ['a', 'e', 'i', 'o', 'u', 'A', 'E', 'I', 'O', 'U']
    for vowel in vowels:
|\codecolorbox{eot}{<EOT>}|
\end{codelisting}
\tcbitem[title={$N=2:$ Left-to-right},colback=white,colframe=bluegray]
\begin{codelisting}{python}
    vowels = ['a', 'e', 'i', 'o', 'u', 'A', 'E', 'I', 'O', 'U']
    for vowel in vowels:
\end{codelisting}
\tcblower
\begin{codelisting}{python}
        text = text.replace(vowel, '')
    return text
\end{codelisting}
\end{tcbitemize}
\caption{An illustration of self-infilling decoding with \textbf{interruption} and \textbf{looping} on \humaneval, which successfully corrects the initially wrong solution at $N=1$ (classified as the `Incorrect $\rightarrow$ Correct' category in \cref{fig:loop_inspect}). The area above the dashed line indicates the current context, while the content below represents the corresponding completion.}
\label{fig:example:loop:wrong_to_correct}
\end{figure}

\begin{figure}[h!]
\begin{DSOuterBox}{Problem Description}
\begin{codelisting}{python}
def sort_array(array):
    """
    Given an array of non-negative integers, return a copy of the given array after sorting,
    you will sort the given array in ascending order if the sum( first index value, last index value) is odd,
    or sort it in descending order if the sum( first index value, last index value) is even.

    Note:
    * don't change the given array.

    Examples:
    * sort_array([]) => []
    * sort_array([5]) => [5]
    * sort_array([2, 4, 3, 0, 1, 5]) => [0, 1, 2, 3, 4, 5]
    * sort_array([2, 4, 3, 0, 1, 5, 6]) => [6, 5, 4, 3, 2, 1, 0]
    """
\end{codelisting}
\end{DSOuterBox}
\begin{tcbitemize}[raster columns=2, raster equal height=rows, size=small]
\tcbitem[title={$N=1:$ Self-infilling},colback=white,colframe=brown]
\tcblower
\begin{codelisting}{python}
    if len(array) == 0:
        return array
    elif len(array) == 1:
        return array
    else:
       |\codecolorbox{suf}{<SUF>}|return array


|\codecolorbox{mid}{<MID>}| if sum(array[0], array[-1]) %
            return sorted(array, reverse=True)
        else:
            |\codecolorbox{eot}{<EOT>}|
\end{codelisting}
\tcbitem[title={$N=1:$ Left-to-right},colback=white,colframe=bluegray]
\begin{codelisting}{python}
    if len(array) == 0:
        return array
    elif len(array) == 1:
        return array
    else:
        if sum(array[0], array[-1]) %
            return sorted(array, reverse=True)
        else:
\end{codelisting}
\tcblower
\begin{codelisting}{python}
            return sorted(array)

\end{codelisting}
\end{tcbitemize}
\begin{tcbitemize}[raster columns=2, raster equal height=rows, size=small]
\tcbitem[title={$N=2:$ Self-infilling},colback=white,colframe=brown]
\begin{codelisting}{python}
|\codecolorbox{suf}{<SUF>}|    else:
        if sum(array[0], array[-1]) %
            return sorted(array, reverse=True)
        else:
            return
\end{codelisting}
\tcblower
\begin{codelisting}{python}
                    sorted(array)

|\codecolorbox{mid}{<MID>}|    if len(array) == 0:
        return []
    elif len(array) == 1:
        return array
|\codecolorbox{eot}{<EOT>}|
\end{codelisting}
\tcbitem[title={$N=2:$ Left-to-right},colback=white,colframe=bluegray]
\begin{codelisting}{python}
    if len(array) == 0:
        return []
    elif len(array) == 1:
        return array
\end{codelisting}
\tcblower
\begin{codelisting}{python}
    else:
        if sum(array[0], array[-1]) %
            return sorted(array, reverse=True)
        else:
            return sorted(array)
\end{codelisting}
\end{tcbitemize}
\caption{An illustration of self-infilling decoding with \textbf{interruption} and \textbf{looping} on \humaneval, where the solution generated at $N=1$ is incorrect and does \emph{not} get fixed after the subsequent update (the `Changed but Remained Incorrect' category in \cref{fig:loop_inspect}). The area above the dashed line indicates the current context, while the content below represents the corresponding completion.}
\label{fig:example:loop:remain_incorrect}
\end{figure}

\begin{figure}[h!]
\begin{DSOuterBox}{Problem Description}
\begin{codelisting}{python}
You are an expert Python programmer, and here is your task: Write a function to find the similar elements from the given two tuple lists. Your code should pass these tests:

assert similar_elements((3, 4, 5, 6),(5, 7, 4, 10)) == (4, 5)
assert similar_elements((1, 2, 3, 4),(5, 4, 3, 7)) == (3, 4)
assert similar_elements((11, 12, 14, 13),(17, 15, 14, 13)) == (13, 14)
[BEGIN]
def similar_elements(test_tup1, test_tup2):
  res = tuple(set(test_tup1) & set(test_tup2))
  return (res) 
[DONE]

... (2 other in-context examples)

You are an expert Python programmer, and here is your task: Write a function to extract the index minimum value record from the given tuples. Your code should pass these tests:

assert index_minimum([('Rash', 143), ('Manjeet', 200), ('Varsha', 100)]) == 'Varsha'
assert index_minimum([('Yash', 185), ('Dawood', 125), ('Sanya', 175)]) == 'Dawood'
assert index_minimum([('Sai', 345), ('Salman', 145), ('Ayesha', 96)]) == 'Ayesha'
[BEGIN]
\end{codelisting}
\end{DSOuterBox}
\begin{tcbitemize}[raster columns=2, raster equal height=rows, size=small]
\tcbitem[title={$N=1:$ Self-infilling},colback=white,colframe=brown]
\tcblower
\begin{codelisting}{python}
def index_minimum(|\codecolorbox{suf}{<SUF>}|return min_value
[DONE]|\codecolorbox{mid}{<MID>}|tuples):
    min_value = min(tuples, key=lambda x: x[1])[0]
    |\codecolorbox{eot}{<EOT>}|
\end{codelisting}
\tcbitem[title={$N=1:$ Left-to-right},colback=white,colframe=bluegray]
\begin{codelisting}{python}
def index_minimum(tuples):
    min_value = min(tuples, key=lambda x: x[1])[0]

\end{codelisting}
\tcblower
\begin{codelisting}{python}
    return min_value
[DONE]
\end{codelisting}
\end{tcbitemize}
\begin{tcbitemize}[raster columns=2, raster equal height=rows, size=small]
\tcbitem[title={$N=2:$ Self-infilling},colback=white,colframe=brown]
\begin{codelisting}{python}
|\codecolorbox{suf}{<SUF>}|    min_value = min(tuples, key=lambda x: x[1])[0]
    return
\end{codelisting}
\tcblower
\begin{codelisting}{python}
           min_value
[DONE]|\codecolorbox{mid}{<MID>}|def index_minimum(tuples):
|\codecolorbox{eot}{<EOT>}|
\end{codelisting}
\tcbitem[title={$N=2:$ Left-to-right},colback=white,colframe=bluegray]
\begin{codelisting}{python}
def index_minimum(tuples):
\end{codelisting}
\tcblower
\begin{codelisting}{python}
    min_value = min(tuples, key=lambda x: x[1])
    return min_value[0]
[DONE]
\end{codelisting}
\end{tcbitemize}
\caption{An illustration of self-infilling decoding with \textbf{interruption} and \textbf{looping} on \mbpp, where the solution generated at $N=1$ remains correct after the subsequent update (the `Changed but Remained Correct' category in \cref{fig:loop_inspect}). The area above the dashed line indicates the current context, while the content below represents the corresponding completion.}
\label{fig:example:loop:remain_correct}
\end{figure}

\begin{figure}[h!]
\begin{DSOuterBox}{Problem Description}
\begin{codelisting}{python}
You are an expert Python programmer, and here is your task: Write a function to find the similar elements from the given two tuple lists. Your code should pass these tests:

assert similar_elements((3, 4, 5, 6),(5, 7, 4, 10)) == (4, 5)
assert similar_elements((1, 2, 3, 4),(5, 4, 3, 7)) == (3, 4)
assert similar_elements((11, 12, 14, 13),(17, 15, 14, 13)) == (13, 14)
[BEGIN]
def similar_elements(test_tup1, test_tup2):
  res = tuple(set(test_tup1) & set(test_tup2))
  return (res) 
[DONE]

... (2 other in-context examples)

You are an expert Python programmer, and here is your task: Write a function to find the maximum difference between the number of 0s and number of 1s in any sub-string of the given binary string. Your code should pass these tests:

assert find_length("11000010001", 11) == 6
assert find_length("10111", 5) == 1
assert find_length("11011101100101", 14) == 2 
[BEGIN]

\end{codelisting}
\end{DSOuterBox}
\begin{tcbitemize}[raster columns=2, raster equal height=rows, size=small]
\tcbitem[title={$N=1:$ Self-infilling},colback=white,colframe=brown]
\tcblower
\begin{codelisting}{python}
def find_length(|\codecolorbox{suf}{<SUF>}|return max_diff
[DONE]|\codecolorbox{mid}{<MID>}|binary_string, n):
  max_diff = 0
  for i in range(n):
    zero_count = 0
    one_count = 0
    for j in range(i, n):
      if binary_string[j] == '0':
        zero_count += 1
      else:
        one_count += 1
      if abs(zero_count - one_count) > max_diff:
        max_diff = abs(zero_count - one_count)
  |\codecolorbox{eot}{<EOT>}|
\end{codelisting}
\tcbitem[title={$N=1:$ Left-to-right},colback=white,colframe=bluegray]
\begin{codelisting}{python}
def find_length(binary_string, n):
  max_diff = 0
  for i in range(n):
    zero_count = 0
    one_count = 0
    for j in range(i, n):
      if binary_string[j] == '0':
        zero_count += 1
      else:
        one_count += 1
      if abs(zero_count - one_count) > max_diff:
        max_diff = abs(zero_count - one_count)

\end{codelisting}
\tcblower
\begin{codelisting}{python}
  return max_diff
[DONE]
\end{codelisting}
\end{tcbitemize}
\begin{tcbitemize}[raster columns=2, raster equal height=rows, size=small]
\tcbitem[title={$N=2:$ Self-infilling},colback=white,colframe=brown]
\begin{codelisting}{python}
|\codecolorbox{suf}{<SUF>}|      if binary_string[j] == '0':
        zero_count += 1
      else:
        one_count += 1
      if abs(zero_count - one_count) > max_diff:
        max_diff = abs(zero_count - one_count)
  return
\end{codelisting}
\tcblower
\begin{codelisting}{python}
          max_diff
[DONE]|\codecolorbox{mid}{<MID>}|def find_length(binary_string, n):
  max_diff = 0
  for i in range(n):
    zero_count = 0
    one_count = 0
    for j in range(i, n):
|\codecolorbox{eot}{<EOT>}|
\end{codelisting}
\tcbitem[title={$N=2:$ Left-to-right},colback=white,colframe=bluegray]
\begin{codelisting}{python}
def find_length(binary_string, n):
  max_diff = 0
  for i in range(n):
    zero_count = 0
    one_count = 0
    for j in range(i, n):

\end{codelisting}
\tcblower
\begin{codelisting}{python}
      if binary_string[j] == '0':
        zero_count += 1
      else:
        one_count += 1
      if abs(zero_count - one_count) > max_diff:
        max_diff = abs(zero_count - one_count)
  return max_diff
[DONE]
\end{codelisting}
\end{tcbitemize}
\caption{An illustration of self-infilling decoding with \textbf{interruption} and \textbf{looping} on \mbpp, where the solution generated at $N=1$ remains unchanged after the subsequent update (the `Unchanged' category in \cref{fig:loop_inspect}). The area above the dashed line indicates the current context, while the content below represents the corresponding completion.}
\label{fig:example:loop:unchanged}
\end{figure}

\begin{figure}[h!]
\begin{DSOuterBox}{Problem Description}
\begin{codelisting}{python}
You are an expert Python programmer, and here is your task: Write a function to find the similar elements from the given two tuple lists. Your code should pass these tests:

assert similar_elements((3, 4, 5, 6),(5, 7, 4, 10)) == (4, 5)
assert similar_elements((1, 2, 3, 4),(5, 4, 3, 7)) == (3, 4)
assert similar_elements((11, 12, 14, 13),(17, 15, 14, 13)) == (13, 14)
[BEGIN]
def similar_elements(test_tup1, test_tup2):
  res = tuple(set(test_tup1) & set(test_tup2))
  return (res) 
[DONE]

... (2 other in-context examples)

You are an expert Python programmer, and here is your task: Write a python function to check whether the frequency of each digit is less than or equal to the digit itself. Your code should pass these tests:

assert validate(1234) == True
assert validate(51241) == False
assert validate(321) == True
[BEGIN]
\end{codelisting}
\end{DSOuterBox}
\begin{tcbitemize}[raster columns=2, raster equal height=rows, size=small]
\tcbitem[title={$N=1:$ Self-infilling},colback=white,colframe=brown]
\tcblower
\begin{codelisting}{python}
def validate(num):
   |\codecolorbox{suf}{<SUF>}|return True
    else:
        return False
[DONE]|\codecolorbox{mid}{<MID>}| num = str(num)
    for i in num:
        if num.count(i) > int(i):
            |\codecolorbox{eot}{<EOT>}|
\end{codelisting}
\tcbitem[title={$N=1:$ Left-to-right},colback=white,colframe=bluegray]
\begin{codelisting}{python}
def validate(num):
    num = str(num)
    for i in num:
        if num.count(i) > int(i):

\end{codelisting}
\tcblower
\begin{codelisting}{python}
            return False
    return True
[DONE]
\end{codelisting}
\end{tcbitemize}
\begin{tcbitemize}[raster columns=2, raster equal height=rows, size=small]
\tcbitem[title={$N=2:$ Self-infilling},colback=white,colframe=brown]
\begin{codelisting}{python}
|\codecolorbox{suf}{<SUF>}|        if num.count(i) > int(i):
            return False
    return
\end{codelisting}
\tcblower
\begin{codelisting}{python}
           True
[DONE]|\codecolorbox{mid}{<MID>}|def validate(num):
    for i in num:
|\codecolorbox{eot}{<EOT>}|
\end{codelisting}
\tcbitem[title={$N=2:$ Left-to-right},colback=white,colframe=bluegray]
\begin{codelisting}{python}
def validate(num):
    for i in num:
\end{codelisting}
\tcblower
\begin{codelisting}{python}
        if num.count(i) > int(i):
            return False
    return True
[DONE]
\end{codelisting}
\end{tcbitemize}
\caption{An illustration of self-infilling decoding with \textbf{interruption} and \textbf{looping} on \mbpp, where the looped update introduces bugs to an initially correct solution at $N=1$ (classified as the `Correct $\rightarrow$ Incorrect' category in \cref{fig:loop_inspect}). The area above the dashed line indicates the current context, while the content below represents the corresponding completion.}
\label{fig:example:loop:correct_to_wrong}
\end{figure}

\end{document}